# An Energy-concentrated Wavelet Transform for Time Frequency Analysis of Transient Signals

Haoran Dong, Gang Yu

*Abstract*—Transient signals are often composed of a series of modes that have multivalued time-dependent instantaneous frequency (IF), which brings challenges to the development of signal processing technology. Fortunately, the group delay (GD) of such signal can be well expressed as a single valued function of frequency. By considering the frequency-domain signal model, we present a postprocessing method called wavelet transform (WT)-based time-reassigned synchrosqueezing transform (WTSST). Our proposed method embeds a two-dimensional GD operator into a synchrosqueezing framework to generate a time-frequency representation (TFR) of transient signal with high energy concentration and allows to retrieve the whole or part of the signal. The theoretical analyses of the WTSST are provided, including the analysis of GD candidate accuracy and signal reconstruction accuracy. Moreover, based on WTSST, the WT-based time-reassigned multisynchrosqueezing transform (WTMSST) is proposed by introducing a stepwise refinement scheme, which further improves the drawback that the WTSST method is unable to deal with strong frequency-varying signal. Simulation and real signal analysis illustrate that the proposed methods have the capacity to appropriately describe the features of transient signals.

## I. Introduction

TRANSIENT signals widely exist in various fields of the real world. With the development of signal detection technology, more and more examples can be acquired in practical applications, such as fault diagnosis of rotating machines, electrocardiogram (ECG) data processing and gravitational wave (GW) signal detection [1-6]. Such signals can usually be accurately modeled as a superposition of amplitude- and frequency-modulated (AM-FM) modes, i.e., multicomponent signal (MCS) [7,8]. Since the change law of the transient signal depends on frequency, the group delay (GD) defined as the derivative of phase to frequency can more effectively reveal the transient features [9,10]. Therefore, it is significant to accurately characterize the GD of transient signal in practical application.

Actually, transient signal is a kind of typical nonstationary signal with frequency-varying characteristics. The classical Fourier transform (FT) considering global frequency is difficult to reflect the characteristic changes of such signal. Time-frequency (TF) analysis (TFA) is one of the effective strategies to locally describe the TF variation features of nonstationary signals, which can be roughly classified into two categories, i.e., linear TFA and quadratic TFA. The typical linear TFA methods, e.g., short-time FT (STFT) and wavelet transform (WT), expand a one-dimensional time series signal into a two-dimensional (2-D) TF plane to observe the complex structure of the signal. They can further obtain the time and frequency information of the signal simultaneously by means of the time-domain translation and frequency-domain modulation of the observation window named Heisenberg box [11-13]. However, conventional TFA methods are limited by the poor energy concentration in the TF plane known as Heisenberg uncertainty principle [14-16]. Moreover, in quadratic TFA methods, Wigner-Ville distribution (WVD) and its variant can obtain TF representation (TFR) with high energy concentration, but introduce undesirable cross term interference to multicomponent signals, which makes it a thorny problem to read TF features clearly [17-20].

Recently, many advanced approaches have been built to refine the resolution of the classical TFA methods, either only in time-domain like empirical mode decomposition (EMD) or in TF plane like reassignment method (RM). As a valuable data-driven scheme, EMD realizes the decomposition of nonstationary multicomponent signal into various intrinsic mode functions (IMFs) and residual function. The instantaneous frequency (IF) of IMF is allowed to be obtained by further applying Hilbert transform (HT). This combination strategy is also called Hilbert-Huang transform (HHT). The properties of HHT and its variants are studied in [21-25]. However, EMD exists an uncontrollable processing behavior due to mode mixing, artifacts and sensitivity to noise, which limits its practical application. And EMD has been unable to find a suitable mathematical foundation to support it [26-28].

The postprocessing schemes of linear TFA have become a hot research field in recent years. Reassignment method (RM) regarded as the focus of postprocessing methods is proposed in [29] and developed in [30]. The main step of RM is to find the location with largest energy, called ridge [31,32], and accumulate TF energy to ridge through two-way reassignment operation, which manages to accurately describe the TF feature. Regretfully, the loss of phase information during reassignment leads to the inherent defect that RM cannot retrieve the original

This work was supported in part by National Natural Science Foundation of China under Grant 61901190. *(Corresponding author: Gang Yu).*

H. Dong and G. Yu are with the School of Electrical Engineering, University of Jinan, Jinan 250022, China (e-mail: dhr19970808@163.com; yugang2010@163.com).

signal. To bridge this gap, a special phase-based RM, namely synchrosqueezing transform (SST) is proposed in [33]. It preserves phase information of original transform by moving TF coefficients instead of reassigning energy. The advantages of SST in reconstructing signal and improving resolution make various SST based schemes developed to enhance the flexibility of SST family [34-37]. In addition, the synchroextracting transform (SET) scheme is proposed in [38], which only retains ridge energy to obtain TFR with better resolution than SST. The theoretical supports of SST and SET are further provided in [33,39] and [40] respectively, which ensures the validity of SST and SET families. Nevertheless, SET and SST based on weak frequency modulated signal framework provide poor TF resolution for strong time-varying signal. To enhance the applicability of SST family, the high-order SST (HSST) and multi-SST (MSST) are given in [41-43] and [44]. HSST obtains the IF candidate closer to the real IF through considering high-order Taylor expansion of signal phase, and then squeezing the coefficients like SST, but the higher order approximation process would bring heavier computational burden. MSST reduces the error remainder between the real IF and IF candidate by the strategy of fixed-point iteration. Meanwhile, the IF in MSST framework only needs second-order approximation model, which greatly reduces the computational burden. However, such frequency direction postprocessing means are merely suitable for coping with signal whose IF is a single valued time-dependent function. Therefore, transient signal with IF as multivalued time-dependent is unable to obtain a readable TFR in SST and SET framework [9,10].

Recently, time-reassigned SST (TSST) and transient transform (TET) have realized the dual operation of SST and SET by considering the frequency-domain signal model and introducing the GD of single valued frequency-dependent function [45,46]. However, TSST and TET are limited to processing weak frequency-varying signals. Moreover, they are embedded under the framework of STFT. Although STFT can improve the shortcomings of FT to some extent, its observation window lacks adaptability. Therefore, STFT cannot be optimized for specific frequency range. Different from STFT, WT allows to flexibly select suitable window length for observing different frequency region of signals, which allows WT can better revealed the feature of signal with singularity points on the appropriate scale [11, 47].

The purpose of this paper is to improve the TSST method by introducing two novel postprocessing methods. We first propose a scheme called WT-based time-reassigned SST (WTSST) for weak frequency-varying signal. We further provide a detailed theoretical analysis of WTSST and analyze the GD error of strong frequency-varying signal under the framework of WTSST. According to that analysis, a novel approach called WT-based time-reassigned MSST (WTMSST) is proposed, which reduces the residual error term of GD candidate to approach the real GD of signal. Moreover, the proposed methods can recover the original signal with a reasonable accuracy.

The remainder of this paper is as follows: in Section II, we review the theory of RM and the definition of energy center of gravity of TFR. Section III provides the theoretical analysis and implementation of the proposed methods, putting the emphasis on the theoretical analysis. The comparison of performance of different TFA methods is illustrated in Section IV. Section V verifies the effectiveness of the proposed methos by dealing with real transient signal. Section VI draws a conclusion.

## II. RELATED WORK

In this paper, we aim to analyze MCS modeled by IMF [10]

$$\hat{x}(\omega) = \sum_{k=1}^{K} A_k(\omega) e^{i\varphi_k(\omega)}, \quad (1)$$

where $A_k(\omega)$ and $\varphi_k(\omega)$ denote the signal amplitude and phase respectively, and the $-\varphi'_k(\omega)$ denotes the GD.

### A. The Modified Wavelet Transform

Traditional linear TFA method links signal $x(t)$ with a waveform dictionary $D = \{\phi_\gamma\}_{\gamma \in \Gamma}$ composed of a series of waveform functions with certain TF locating ability, i.e., $TFA(\tau, \omega) = \langle x(t), \phi_\gamma(t) \rangle$, where $\langle,\rangle$ and $\gamma$ denote the inner product operator and a multi-index parameter set respectively. The result of the inner product indicates that the position and size of the observation area named Heisenberg box depend on the TF center and span of $\phi_\gamma(t)$. When the TF index changes in the $\mathbb{R}^2$, the Heisenberg box covers the entire TF plane [11].

A WT atom is modelled by a scale factor $a$ and translation factor $b$, i.e., $\phi_\gamma(t) = \psi_{a,b}(t) = 1/a \, \psi((t-b)/a)$. The resulting WT of $x(t) \in L^2(\mathbb{R})$ is

$$W(a,b) = \frac{1}{a} \int_{-\infty}^{+\infty} x(t) \psi^*((t-b)/a) dt \quad (2)$$

where $(\ )^*$ denotes the complex conjugation. We suppose that the wavelet is analytic, which means that it can be defined as a real window function $g(t)$ modulated by the $\omega_0$, i.e., $\psi(t) = g(t)e^{i\omega_0 t}$, then we have

$$W(a,b) = \frac{1}{a} \int_{-\infty}^{+\infty} x(t) \left( g((t-b)/a) e^{i\omega_0 (t-b)/a} \right)^* dt \quad (3)$$

where $g^* = g$. When the regular WT expression applies an extra phase shift $e^{-i\omega_0 b/a}$, the function of modified WT (MWT) is defined as

$$W_e(a,b) = \frac{1}{a} \int_{-\infty}^{+\infty} x(t) \left( g((t-b)/a) e^{i\omega_0 t/a} \right)^* dt. \quad (4)$$

Letting $g_{\omega_0}(t) = g((t-b)/a) e^{i\omega_0 t/a}$ yields

$$W_e(a,b) = \frac{1}{a} \int_{-\infty}^{+\infty} x(t) g_{\omega_0}^*(t) dt$$
$$= \frac{1}{2\pi a} \int_0^{+\infty} \hat{x}(\xi) \hat{g}_{\omega_0}^*(\xi) d\xi. \quad (5)$$

Consider $\hat{g}_{\omega_0}(\xi)$ to be calculated by

$$\hat{g}_{\omega_0}(\xi) = \int_{-\infty}^{+\infty} g((t-b)/a) e^{i\omega_0 t/a} e^{-i\xi t} dt. \quad (6)$$

It is worth noting that there is a conversion relationship between

$\omega$ and $a$: $\omega = \omega_0/a$. Putting it into (6) and letting $u = (t-b)/a$ yield

$$\hat{g}_{\omega_0}(\xi) = ae^{-i(\xi-\omega)b}\int_{-\infty}^{+\infty} g(u)e^{-ia(\xi-\omega)u}du \\ = a\hat{g}(a(\xi-\omega))e^{-i(\xi-\omega)b}. \quad (7)$$

Substituting (7) into (5) obtains

$$W_e(a,b) = \frac{1}{2\pi}\int_0^{+\infty} \hat{x}(\xi)\hat{g}^*(a(\xi-\omega))e^{i(\xi-\omega)b}d\xi. \quad (8)$$

### B. The Reassignment Candidates

Considering that the serious diffusion of MWT energy is caused by the Heisenberg box, we need to explore the energy distribution law of the MWT. According to Plancherel's theorem, the spectrogram $P(a,b)$ of MWT can be described as follows

$$P(a,b) = |W_e(a,b)|^2 = W_e(a,b)\overline{W_e(a,b)} \\ = \frac{1}{2\pi}\int x(t)g\left(\frac{t-b}{a}\right)e^{-i\omega t}dt\int \hat{x}(\xi)\hat{g}^*(a(\xi-\omega))e^{i(\xi-\omega)b}d\xi \\ = \iint \frac{1}{2\pi}x(t)\hat{x}^*(\xi)g\left(\frac{t-b}{a}\right)\hat{g}(a(\xi-\omega))e^{-i\omega t}e^{-i(\xi-\omega)b}dtd\xi \\ = \iint Q_{a,b}(t,\xi)dtd\xi, \quad (9)$$

where $Q_{a,b}(t,\xi)$ denotes the probability distribution function around the point $(a,b)$ [46,48]. Therefore, the center of gravity of signal energy under the MWT can be defined by

$$\hat{t}(a,b) = \frac{\iint tQ_{a,b}(t,\xi)dtd\xi}{\iint Q_{a,b}(t,\xi)dtd\xi} \\ = \frac{\int tx(t)\hat{g}^*((t-b)/a)e^{-i\omega t}dt}{\int x(t)\hat{g}^*((t-b)/a)e^{-i\omega t}dt}, \quad (10)$$

$$\hat{\omega}(a,b) = \frac{\iint \xi Q_{a,b}(t,\xi)dtd\xi}{\iint Q_{a,b}(t,\xi)dtd\xi} \\ = \frac{\int \xi \hat{x}(\xi)\hat{g}^*(a(\xi-\omega))e^{i(\xi-\omega)b}d\xi}{\int \hat{x}(\xi)\hat{g}^*(a(\xi-\omega))e^{i(\xi-\omega)b}d\xi} \\ = \omega + \frac{\partial_b W_e(a,b)}{iW_e(a,b)} = \omega + \frac{W_e^{\xi g}(a,b)}{aW_e(a,b)}. \quad (11)$$

The detailed derivations are provided in Appendix A. The aim of RM is to reassign the points that deviate from the center of the gravity of signal energy by considering new TF point $(\hat{\omega}(a,b),\hat{t}(a,b))$ instead of $(a,b)$, i.e.,

$$R(\eta,\tau) = \iint P(a,b)\delta(\tau-\hat{t}(a,b))\delta(\eta-\hat{\omega}(a,b))dadb. \quad (12)$$

RM greatly enhances the energy concentration of linear TFA, but its bidirectional reassignment scheme hinders its ability to reconstruct the original signal.

## III. WT-BASED TIME-REASSIGNED SYNCHROSQUEEZING TRANSFORM

### A. Time Reassignment of Wavelet Transform

Due to the limitation of the Heisenberg box area, the resolution of the traditional linear TFA method cannot describe the transient events accurately. The numerical transient signal containing two modes is constructed in Fig. 1(a) and the corresponding TFRs implemented by MWT and STFT are shown in Fig. 1(b) and 1(c) to intuitively reflect the energy diffusion phenomenon.

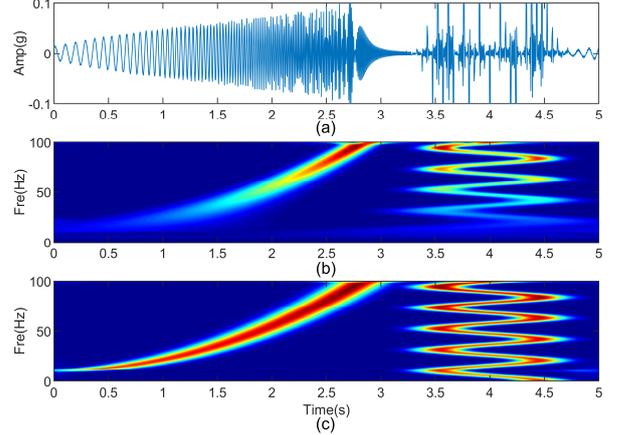

**Fig. 1.** (a) Signal waveform, (b) WT result and (c) STFT result.

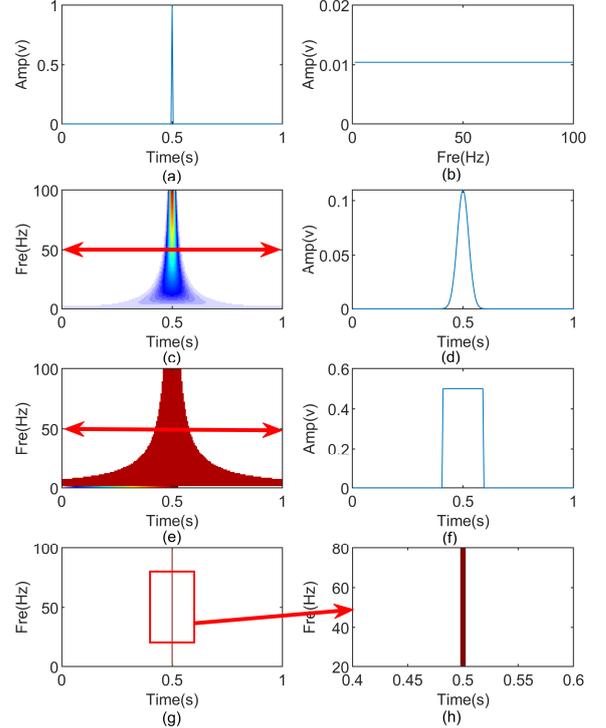

**Fig. 2.** (a) The Dirac function signal with $t_0 = 0.5s$, (b) the frequency-domain representation, (c) the MWT spectrogram, (d) the slice of the MWT spectrogram at $\omega_1 = 50Hz$, i.e., $|W_e(\omega_0/\omega_1,b)|$, (e) the GD $\hat{t}(a,b)$, (f) the slice of the GD at $\omega_1 = 50Hz$, i.e., $\hat{t}(\omega_0/\omega_1,b)$, (g) the WTSST result and (h) the zoomed version of WTSST result

To enhance the energy concentration of MWT and retain the reconstruction ability, we only consider the postprocessing strategy along the time direction. Thus, the WTSST can be

defined by the operator $\int \delta(u - \hat{t}(a,b))db$

$$S(a,u) = \int_{A(\omega)} W_e(a,b)\delta(u - \hat{t}(a,b))db, \quad (13)$$

where $A(\omega) = \{b; W_{e,x}^g(\omega_0/\omega, b) \neq 0\}$. The nature of WTSST is the accumulation process by transferring the time-scale (TS) coefficients from the point $(a,b)$ to the new point $(a, \hat{t}(a,b))$. More importantly, WTSST retains the ability to recover original signal i.e.,

$$\hat{x}(\omega) = \frac{1}{\hat{g}(0)} \int_{-\infty}^{+\infty} S(a,u) du. \quad (14)$$

Herein, we employ a discrete Dirac digital signal $x(t) = \delta(t - 0.5)$ with a sampling frequency of 200Hz to illustrate the WTSST. Substituting $x(t)$ into (4), the resulting MWT is

$$W_e(a,b) = Ag((t_0 - b)/a)e^{-i\omega_0 t_0/a}/a. \quad (15)$$

The time-domain precise locating capacity and undesired frequency bandwidth of the Dirac signal are revealed in Fig. 2(a) and 2(b), respectively. It can be observed in Fig. 2(c) that in the time direction, the TFR, which should be as ideal as the Fig. 2(a), forms a certain diffusion region near the trajectory $t_0 = 0.5$. According to (15), the time-domain compactly supported window $g(\bullet)$ implies that the energy $|W_e(a,b)|$ is distributed around the $t = t_0$ and reaches the maximum at that time. To explain intuitively, a slice of the MWT spectrogram at the frequency point $\omega = 50Hz$ is depicted in Fig. 2(d). It illustrates that the window length $[t_0 - ad, t_0 + ad]$ affects the energy distribution area of the MWT, where $d$ denotes time-domain compactly support of the basis window. Therefore, the most important issue in linear TFA is the Heisenberg uncertainty principle that restricts the energy concentration ability. Combining (10) and (15), it can be proved that when $W_e(a,b) \neq 0$, the position of the center of gravity of Dirac signal energy is consistent with its time ideal position, i.e.,

$$\hat{t}(a,b) = \frac{t_0 Ag((t_0-b)/a)e^{-i\omega_0 t_0/a}/a}{Ag((t_0-b)/a)e^{-i\omega_0 t_0/a}/a} = t_0. \quad (16)$$

We can also find from Fig. 2(e) and 2(f) that, in the $t \in [t_0 - \Delta, t_0 + \Delta]$ where $\Delta = ad$, all values of the 2-D GD are equal to 0.5. Further employing (13) leads to a TFR with a significantly enhanced TF resolution in Fig. 2(g) and 2(h).

*B. WTSST Performance Analysis*

Since WTSST can obtain theoretically guaranteed accuracy when processing weak frequency-varying signals, we define the following IMF that meets the WTSST framework to provide the theoretical support.

**Remark:** The signal is easily affected by noise and becomes unstable at $|W_{e,x}^g(a,b)| = 0$. Therefore, it is necessary to consider a threshold-limited region for $\hat{t}(a,b)$ to eliminate such undefined points, which is equivalent to replacing $A(\omega)$ with the smaller region i.e., $A_\varepsilon(\omega) := \{b; |W_{e,x}^{\hat{g}}(\omega_0/\omega, b)| \geq \varepsilon\}$.

**Definition 1:** Let the class $A_{\varepsilon,d}$ represent a set of all superpositions of well-separated IMFs satisfying threshold $\varepsilon > 0$ and separation distance $\Delta = ad > 0$, if the signal $\hat{x}(\omega) = \sum_{k=1}^{K} \hat{x}_k(\omega) = \sum_{k=1}^{K} A_k(\omega)e^{i\varphi_k(\omega)} \in L^\infty(\mathbb{R})$ can present the following two properties:

1) $A_k(\omega)$ and $\varphi_k(\omega)$ satisfy:

$A(\omega) \in C^1(\mathbb{R}) \cap L^\infty(\mathbb{R}), \quad \varphi(\omega) \in C^2(\mathbb{R}),$

$\inf_{t \in \mathbb{R}} \varphi'(\omega) > 0, \quad \sup_{t \in \mathbb{R}} \varphi'(\omega) < \infty,$

$|A'(\omega)|, |\varphi''(\omega)| \leq \varepsilon|\varphi'(\omega)|, \quad \forall \omega \in \mathbb{R}^+,$

$M'' := \sup|\varphi''(\omega)| < \infty.$

2) functions $\hat{x}_k$ present obvious separation distance $d > 0$:

$\varphi'_k(\omega) > \varphi'_{k-1}(\omega)$ and $\varphi'_k(\omega) - \varphi'_{k-1}(\omega) \geq 2\Delta = 2ad, \forall \omega \in (\mathbb{R}^+)$

Definition 1 denotes that signal $\hat{x} \in A_{\varepsilon,d}$ under the WTSST framework is composed of a series of components with slowly frequency-varying amplitude and sufficiently smooth delay, and there is no contact between these components. The main result of WTSST is illustrated by the following theorem.

**Theorem 1.** Let $\hat{x}(\omega) \in A_{\varepsilon,d}$ and set $\tilde{\varepsilon} = \varepsilon^{1/3}$. Select a function $\hat{h} \in C^\infty(\mathbb{R})$ with $\int \hat{h}(\omega)d\omega = 1$ and a window $\hat{g} \in S$ in Schwartz class such that its time-domain form $g$ is compactly supported in $[-\Delta, \Delta]$ where $\Delta = ad$. The WTSST result of $\hat{x}(\omega) \in A_{\varepsilon,d}$ with the threshold $\tilde{\varepsilon}$ and accuracy $\delta$ can be defined by

$$S_{x,\tilde{\varepsilon}}^\delta(a,u) = \int_{A_{\tilde{\varepsilon},x}(\omega)} W_{e,x}^{\hat{g}}(a,b) \frac{1}{\delta} \hat{h}\left(\frac{u - \hat{t}(a,b)}{\delta}\right) db,$$

where $A_{\tilde{\varepsilon},x}(\omega) := \{b \in \mathbb{R}; |W_{e,x}^{\hat{g}}(\omega_0/\omega, b)| > \tilde{\varepsilon}\}$ and the $\varepsilon$ is small enough. Then, the following holds:

1) $|W_{e,x}^{\hat{g}}(a,b)| > \tilde{\varepsilon}$ only when, for each $k \in \{1,\ldots,K\}$, $(a,b) \in Z_k := \{(a,b); |b + \varphi'(\omega)| < \Delta\}$.

2) for each $k \in \{1,\ldots,K\}$ and each $(a,b) \in Z_k$, such that $|W_{e,x}^{\hat{g}}(a,b)| > \tilde{\varepsilon}$, we have $|\hat{t}(a,b) + \varphi'(\omega)| \leq \tilde{\varepsilon}$.

The detailed proofs in Theorem 1 are provided in Appendix B. Theorem 1 indicates that, for $\hat{x}(\omega) \in A_{\varepsilon,d}$, if the selected window meets the band limited condition, the TF distribution of non-zero coefficients is concentrated in a narrow band near the trajectory $(a, -\varphi'_k(\omega_0/a))$.

This subsection shows that WTSST is indeed a reliable tool when coping with weak frequency-varying signals. However, transient signals in reality always present GD with more complex modulation laws, which may make WTSST no longer have high resolution. The next subsection will analyze the performance of WTSST in the case of complex frequency-varying GD.

*C. Accuracy analysis of WTSST under Second-order GD*

**Definition 2.** The class $B_{\varepsilon,d}$ denotes a set of all superpositions of well-separated IMFs satisfying threshold $\varepsilon > 0$ and separation distance $\Delta > 0$, where the signal $\hat{x}(\omega) \in L^{\infty}(\mathbb{R})$ keeps the following two hypotheses:

1) $A_k(\omega)$ and $\varphi_k(\omega)$ satisfy the following conditions:
$A_k(\omega) \in C^1(\mathbb{R}) \cap L^{\infty}(\mathbb{R})$, $\varphi_k(\omega) \in C^2(\mathbb{R})$,
$\inf_{t \in \mathbb{R}} \varphi'_k(\omega) > 0$, $\sup_{t \in \mathbb{R}} \varphi'_k(\omega) < \infty$,
$|A'_k(\omega)|, |A''_k(\omega)|, |\varphi'''_k(\omega)| \leq \varepsilon |\varphi'_k(\omega)|$, $\forall \omega \in \mathbb{R}^+$,
$M := \sup |A_k(\omega)| < \infty, M' := \max(\sup |A'_k(\omega)|, \sup |\varphi'_k(\omega)|)$,
$M'' := \sup |\varphi''_k(\omega)| < \infty$, $M''' := \sup |\varphi'''_k(\omega)| < \infty$.

2) function $\hat{x}_k$ has obvious separation distance $d > 0$:
$\varphi'_k(\omega) > \varphi'_{k-1}(\omega)$ and $\varphi'_k(\omega) - \varphi'_{k-1}(\omega) \geq 2\Delta = 2ad$, $\forall \omega \in (\mathbb{R}^+)$.

We consider a signal with GD that can be expanded by second-order Taylor, i.e., $\varphi(\omega) = \sum_{k=0}^{2} \omega^k \beta_k / k!$ where $\beta_k \in \mathbb{R}$. The resulting MWT can be obtained as

$$W_e(a,b) = \int_{\mathbb{R}^+} \hat{x}(\xi + \omega) \hat{g}(a\xi) e^{i\xi b} d\xi$$
$$= A(\omega) \int_{\mathbb{R}^+} \exp\left(\sum_{k=0}^{2} \frac{i}{k!} \beta_k (\xi + \omega)^k\right) \quad (17)$$
$$\times \hat{g}(a\xi) e^{i\xi b} d\xi.$$

Furthermore, according to Parseval's theorem, Equation (10) can be rewritten as

$$\hat{t}(a,b) = \frac{\int t x(t) \hat{g}((t-b)/a) e^{-i\omega_0 t/a} dt}{\int x(t) \hat{g}((t-b)/a) e^{-i\omega_0 t/a} dt}$$
$$= \frac{i \int \hat{x}'(\xi) \hat{g}^*(a(\xi - \omega)) e^{i(\xi - \omega_0/a)b} d\xi}{\int \hat{x}(\xi) \hat{g}^*(a(\xi - \omega)) e^{i(\xi - \omega_0/a)b} d\xi}. \quad (18)$$

Combing (17) and (18), the local GD candidate can be obtained for $|W_{e,x}^{\hat{g}}| \neq 0$

$$\hat{t}(a,b) = \sum_{k=1}^{2} (-\beta_k) \omega^{k-1} + (-\beta_2) \frac{W_e^{\xi \hat{g}}}{a W_e^{\hat{g}}} \quad (19)$$

where $-\varphi'(\omega) = -\beta_1 - \beta_2 \omega$ and $-\varphi''(\omega) = \beta_2$. For the convenience of observing the error of the GD candidate, we choose the Gaussian window $\hat{g}(\omega) = \sqrt{2\pi}\sigma e^{-0.5\sigma \omega^2}$ to specify (19)

$$\hat{t}(a,b) = -\varphi'(\omega) - \varphi''(\omega) \frac{W_e^{\xi \hat{g}}}{a W_e^{\hat{g}}}$$
$$= -\varphi'(\omega) + \frac{\varphi''(\omega)^2}{\varphi''(\omega)^2 + (a^2 \sigma)^2} (\varphi'(\omega) + b). \quad (20)$$

The specific derivation of (20) is described in Appendix C. Equation (20) means that the $\hat{t}(a,b)$ is failed to accurately estimate the $-\varphi'(\omega)$ of the signal with strong frequency-varying GD.

*D. Fixed Point Iteration Scheme*

In this subsection, we introduce the fixed-point iteration strategy to compensate for the error between $\hat{t}(a,b)$ and $-\varphi'(\omega)$ in (20). Within this idea, we propose the WTMSST method and formulate it as

$$S^{[2]}(a,u) = \int_{-\infty}^{+\infty} S^{[1]}(a,b) \delta(u - \hat{t}(a,b)) db$$
$$S^{[3]}(a,u) = \int_{-\infty}^{+\infty} S^{[2]}(a,b) \delta(u - \hat{t}(a,b)) db \quad (21)$$
$$\vdots$$
$$S^{[N]}(a,u) = \int_{-\infty}^{+\infty} S^{[N-1]}(a,b) \delta(u - \hat{t}(a,b)) db,$$

where $S(a,u)$ in (13) is represented here as $S^{[1]}(a,b)$ and $N$ is the iteration number such that $N \geq 2$.

In order to obtain the law in the process of multiple iterations, it is necessary to analyze the relationship between WTMSST ($N = 2$) and WTSST. Substituting $S^{[1]}(a,b)$ into $S^{[2]}(a,b)$ and combining Fubini's theorem yield

$$S^{[2]}(a,u) = \int S^{[1]}(a,\tau) \delta(u - \hat{t}(a,\tau)) d\tau$$
$$= \iint W_e(a,b) \delta(\tau - \hat{t}(a,b)) db$$
$$\times \delta(u - \hat{t}(a,\tau)) d\tau$$
$$= \int W_e(a,b) \int \delta(\tau - \hat{t}(a,b)) \quad (22)$$
$$\times \delta(u - \hat{t}(a,\tau)) d\tau db$$
$$= \int W_e(a,b) \delta(u - \hat{t}(a,\hat{t}(a,b))) db.$$

Equation (22) illustrates that the iterative operation of WTSST is equivalent to squeezing the WT coefficients into a newly generated GD candidate $\hat{t}(a,\hat{t}(a,b))$. Substituting $S^{[2]}(a,b)$ into $S^{[3]}(a,b)$ allows the iterative law to be more obviously demonstrated:

$$S^{[3]}(a,u) = \int S^{[2]}(a,\tau) \delta(u - \hat{t}(a,\tau)) d\tau$$
$$= \iint W_e(a,b) \delta(\tau - \hat{t}(a,\hat{t}(a,b))) db \quad (23)$$
$$\times \delta(u - \hat{t}(a,\tau)) d\tau$$
$$= \int W_e(a,b) \delta(u - \hat{t}(a,\hat{t}(a,\hat{t}(a,b)))) db.$$

Compared with $N = 2$, $N = 3$ produces a new operator $\hat{t}(a,\hat{t}(a,\hat{t}(a,b)))$ again. If the iteration step continues, we can always update the GD candidate. We further introduce $\hat{t}^{[N]}(a,b)$ to replace the original GD estimator $\hat{t}(a,b)$ in WTSST, and then we can get the expression of the WTMSST as

$$S^{[N]}(a,u) = \int_{-\infty}^{+\infty} W_e(a,b) \delta(u - \hat{t}^{[N]}(a,b)) db. \quad (24)$$

According to (20), we can describe the $\hat{t}^{[N]}(a,b)$ as

$$\hat{t}^{[N]}(a,b) = -\varphi'(\omega) + \left[\frac{\varphi''(\omega)^2}{\varphi''(\omega)^2 + (a^2 \sigma)^2}\right]^N (\varphi'(\omega) + b). \quad (25)$$

When the number of iterations $N$ is large enough, the error between the new GD candidate $\hat{t}^{[N]}(a,b)$ and $-\varphi'(\omega)$ would close in 0, i.e., $\lim_{N \to \infty} \hat{t}^{[N]}(a,b) = -\varphi'(\omega)$. It proves that the new operator is more suitable for handling strong frequency-varying signal than $\hat{t}(a,b)$. Furthermore, WTMSST can accurately locate the energy ridges, i.e.,

$$\lim_{N \to \infty} S^{[N]}(a,u) = \int W_e(a,b) \delta(u + \varphi'(\omega)) db$$

$$= \frac{1}{2\pi} \iint \hat{x}(\xi) \hat{g}^*(a(\xi-\omega)) e^{i(\xi-\omega)b} d\xi \quad (26)$$
$$\times \delta(u + \varphi'(\omega)) db$$
$$= \hat{x}(\omega) \hat{g}(0) \delta(u + \varphi'(\omega)).$$

To intuitively reflect the improvement effect of WTMSST on WTSST, we provide their TFRs for processing the signal in Fig. 1, respectively [see Fig. 3]. The results show that both the WTSST and WTMSST are more appropriate for treating transient signals than WT and STFT. However, it can be seen from the zoomed version that for signals with nonlinear GD, the TFR of WTSST is still blurred whereas the WTMSST allows to provide a sharpened TFR. Moreover, the WTMSST not only enhances the resolution of WTSST, but also retains its reconstruction ability, i.e.,

$$\hat{x}(\omega) = \frac{1}{\hat{g}(0)} \int_{-\infty}^{+\infty} S^{[N]}(a,u) du, \quad (27)$$

$$x(t) = \frac{1}{2\pi \hat{g}(0)} \int_0^{+\infty} \int_{-\infty}^{+\infty} S^{[N]}(a,u) \frac{\omega_0}{a^2} e^{i\frac{\omega_0}{a}t} du da. \quad (28)$$

It is worth noting that the WTMSST strategy has the same reconstruction expression as WTSST. In the next subsection, we will provide the reconstruction accuracy analysis of both.

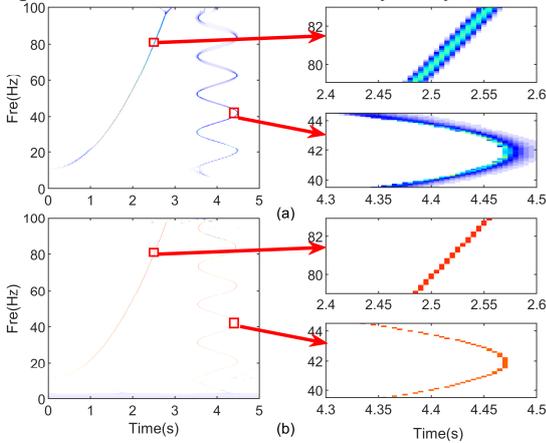

Fig. 3. The WTSST and WTMSST representation of the numerical signal in Fig.1.

*E. Theoretical Analysis of Reconstruction Accuracy*

Both WTSST and WTMSST allow the recovery of an original signal with a reasonable accuracy when Definition 1 and 2 are satisfied respectively. Therefore, there exists a constant C such that, for each $k \in \{1,\ldots,K\}$ and any $\omega \in \mathbb{R}^+$

$$\left| \lim_{\delta \to 0} \left( \frac{1}{\hat{g}(0)} \int_{|u+\varphi_k'(\omega)|<\tilde{\varepsilon}} S_{x,\tilde{\varepsilon}}^{\delta}(a,u) du \right) - A_k(\omega) e^{i\varphi_k(\omega)} \right| \leq C\tilde{\varepsilon},$$

$$\left| \lim_{\delta \to 0} \left( \frac{1}{\hat{g}(0)} \int_{|u+\varphi_k'(\omega)|<\tilde{\varepsilon}} S_{x,\tilde{\varepsilon}}^{[N],\delta}(a,u) du \right) - A_k(\omega) e^{i\varphi_k(\omega)} \right| \leq C\tilde{\varepsilon}.$$

It means that the mode in the kth narrow band can be reconstructed from WTSST and WTMSST with high precision.

The detailed reconstruction error analyses are shown in Appendix D.

*F. Algorithm implementation*

The discretization of the proposed methods is implemented in this subsection to enhance their practicability. The signals observed in the real world are both discrete and real-valued, i.e., $x[n], n = 0,1,\ldots,L-1$ where $L$ is the signal length. $x[n]$ is regarded as a uniform discretization of $x(t)$ taken at the time instant $t_n = t_0 + nT$, where $T$ is the sampling interval. Therefore, the angular frequency interval is $\Delta\xi = 2\pi(LT)^{-1}$ and the discrete version is $\xi[k] = k\Delta\xi$.

The first step is to achieve the discretization of the MWT. $W_e[k,n], n,k = 0,1,\ldots,L-1$ is used to replace the discretized of $W_e(a,b)$

$$W_e[k,n] = \sum_{j=0}^{L-1} x[j] \psi_k^*[j-n] e^{-i\omega_0 n/a_k}, \quad (29)$$

where $\psi_k[\cdot]$ is the N-point wavelet function, i.e.,

$$\psi_k[n] = \frac{1}{a_k} g\left(\frac{n}{a_k}\right) e^{i\omega_0 n/a_k}, \quad (30)$$

where $a_k$ are discrete scales.

The second step is the acquisition of the discretized GD. In practice, we only retain GD with "large enough" TF coefficients and ignore the points with small coefficients. When $|W_e^g| \approx 0$, the GD is unstable, and the signal has been contaminated by noise. Therefore. we define the following numerical support to acquire GD:

$$S_\Upsilon[k] = \{n : |W_e^g[k,n]| > \Upsilon\}, \text{ for } k = 0,1,\ldots,L-1, \quad (31)$$

where the hard threshold $\Upsilon > 0$ is selected to counteract unstable situations. For the MWT, the following expression is used to calculate the GD:

$$\hat{t}(a,b) = \Re\left\{ \frac{\int tx(t) g^*((t-b)/a) e^{-i\omega_0 t/a} dt}{\int x(t) g^*((t-b)/a) e^{-i\omega_0 t/a} dt} \right\}$$
$$= b + \Re\left\{ \frac{aW_e^{tg}(a,b)}{W_e^g(a,b)} \right\}. \quad (32)$$

We can use $W_e^{tg}[k,n]$ instead of $W_e^{tg}(a,b)$. Therefore, for $n \in S_\Upsilon[k]$, we have:

$$\hat{t}[k,n] = n + \Im\left\{ \frac{a_k W_e^{tg}[k,n]}{W_e^g[k,n]} \right\}. \quad (33)$$

The final step in WTSST is to accumulate the MWT coefficients to the same GD position on the TS plane. For the discrete signal $x[n]$, the sum of different contributions has the same effect as the synchrosqueezing process

$$S[k,\tau] = \sum_{n=-0}^{L-1} W_e[k,n] \delta[\tau - \hat{t}[k,n]], \quad (34)$$

where the $S[k,\tau]$ depends on the center $t_l$ of a series of durations $[t_l-\Delta, t_l+\Delta]$ with $t_l-t_{l-1}=2\Delta$. The pseudo code of WTSST is described as

**Algorithm 1** Implementation of WTSST

**Step 1 : Initialization**
Choose the wavelet $\psi$ ;
Choose the threshold $\Upsilon$ ;
Calculate the discrete wavelet $\psi, tg$;
$S[k,n] \leftarrow 0$;
**Step 2 : Calculate the WT representation**
Calculate $W_e[k,n]$ in (29);
**Step 3 : Calculate the GD estimator**
Choose $S_\Upsilon[k]$ ;
Calculate $\hat{t}[k,n]$ in (33);
**Step 4 : Synchrosqueezing**
for $k=1:L$
  for $n=1:L$
    if $n \in S_\Upsilon[k]$
      $\tau \leftarrow \hat{t}[k,n]$;
      $S[k,\tau] \leftarrow S[k,\tau] + W_e[k,n]$;
    end if
  end for
end for
**output** $S[k,\tau]$

Actually, the GD candidate of WTMSST has two convergence modes, which are listed in Appendix E. According to the (26), the discrete version of the WTMSST can be written as

$$S^{[N]}[k,\tau] = \sum_{n=0}^{L-1} W_e[k,n]\delta\left[\tau - \hat{t}^{[N]}[t,n]\right], \quad (35)$$

where the acquisition of $\hat{t}^{[N]}[t,n]$ includes linear iteration and exponential iteration. The pseudo-code for linear iteration is shown as

**Algorithm** 2 Implementation of linear WTMSST

**Step 1 : Initialization**
Choose the wavelet $\psi$ and iteration number $N$;
Choose the threshold $\Upsilon$;
Calculate the discrete wavlet $\psi$ $tg$;
**Step 2 : Calculate the WT representation**
Calculate $W_e[k,n]$ in (29);
**Step 3 : Calculate the GD estimator**
Choose $S_\Upsilon[k]$;
Calculate $\hat{t}[k,n]$ in (33);
**Step 4 : Synchrosqueezing**
$S^{[1]}[k,n] \leftarrow 0$;
$S^{[0]}[k,n] \leftarrow W_e[k,n]$;
for $m=1:N$
  for $k=1:L$
    for n $=1:L$
      $\tau \leftarrow \hat{t}[k,n]$;
      $S^{[m]}[k,\tau] \leftarrow S^{[m]}[k,\tau] + S^{[m-1]}[k,\tau]$;
    end for
  end for
end for
**output** $S^{[N]}[k,\tau]$

The exponential iteration version can be obtained as

**Algorithm** 2 Implementation of exponential WTMSST

**Step 1** : **Initialization**
Choose the wavelet $\psi$ and iteration number $2^N$;
Choose the threshold $\Upsilon$;
Calculate the discrete wavlet $\psi$ $tg$;
**Step 2 : Calculate the WT representation**
Calculate $W_e[k,n]$ in (29);
**Step 3 : Calculate the GD estimator**
Choose $S_\Upsilon[k]$;
Calculate $\hat{t}[k,n]$ in (33);
**Step 4 : Fixd point iteration**
$\hat{t}^{[0]}[k,n] \leftarrow 0; \hat{t}^{[1]}[k,n] = \hat{t}[k,n]$;
for $m=2^{0:N}$
  for $k=1:L$
    for $n=1:L$;
      $\tau \leftarrow \hat{t}^{[m/2]}[k,n]$;
      $\hat{t}^{[m]}[k,n] \leftarrow \hat{t}^{[m/2]}[k,\tau]$;
    end for
  end for
end for
**Step 5 : Synchrosqueezing**
$S^{[2^N]}[k,n] \leftarrow 0$;
for $n=1:L$
  for $k=1:L$
    if $n \in S_\Upsilon[k]$;
      $\tau \leftarrow \hat{t}^{[2^N]}[k,n]$;
      $S^{[2^N]}[k,\tau] \leftarrow S^{[2^N]}[k,\tau] + W_e[k,n]$;
    end if
  end for
end for
**output** $S^{[2^N]}[k,\tau]$

The above pseudo-code of WTMSST and Appendix E illustrate that the convergence rates of the two methods are $N$ and $\log_2 N$ respectively. When $N$ tends to infinity, the convergence rate of nonlinear iteration is faster than that of the liner, i.e., $\lim_{N\to\infty} \log_2 N / N = 0$.

## IV. NUMERICAL VALIDATION

The introduction of objective quantitative index can help us better evaluate the performance of the proposed methods. The aggregation degree of information content can usually be correlated with the fluctuation of information entropy. Therefore, Rényi entropy is an appropriate indicator to characterize the energy concentration of TFA result. The more concentrated the energy of TFR becomes, the stronger its positioning ability is and the TFA methods with lower Rényi entropy usually correspond to the better concentration. In order to further evaluate the robustness of the WTSST and WTMSST, several advanced TFA methods are applied to test the signal in Fig. 1(a) doped with Gaussian white noise with SNR from 1 to 30 dB. The Rényi entropies of different TFA tools are depicted in Fig. 4. It can be seen that with the increase of noise, the entropy values also increase, which denotes that the presence of noise will reduce the energy concentration of all TFA methods. However, the entropies of WTSST and WTMSST, especially WTMSST, are smaller than that of other methods under any SNR, which means that when noise contamination deteriorates, our proposed methods still have the capacity to accurately describe the transient events.

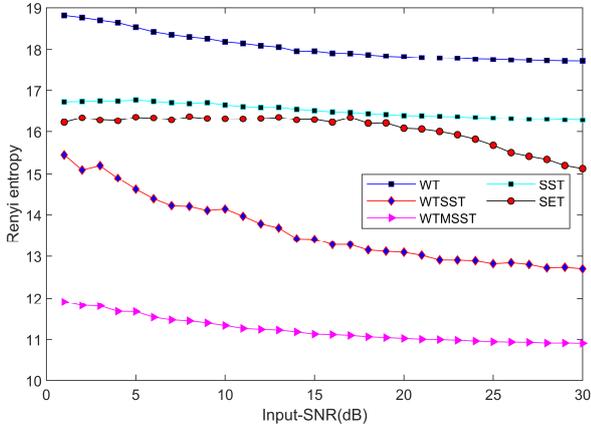

**Fig. 4.** The Rényi entropies of the TF results by different TFA methods under different noise levels (SNR of 1 dB-30 dB).

We show the TFRs of different processing tools when the SNR is equal to 10 dB. Due to the restriction of the Heisenberg uncertainty principle and the interference of noise, the energy diffusion of WT in Fig. 5(a) is serious. When dealing with such numerical signals with strong amplitude modulation, SST and SET are unable to appropriately reassign the blurred energy in the right direction [see Fig. 5(d) and 5(e)]. Fortunately, the WTSST and WTMSST results in Fig. 5 (b) and 5(c) show an appropriate sharpened representation.

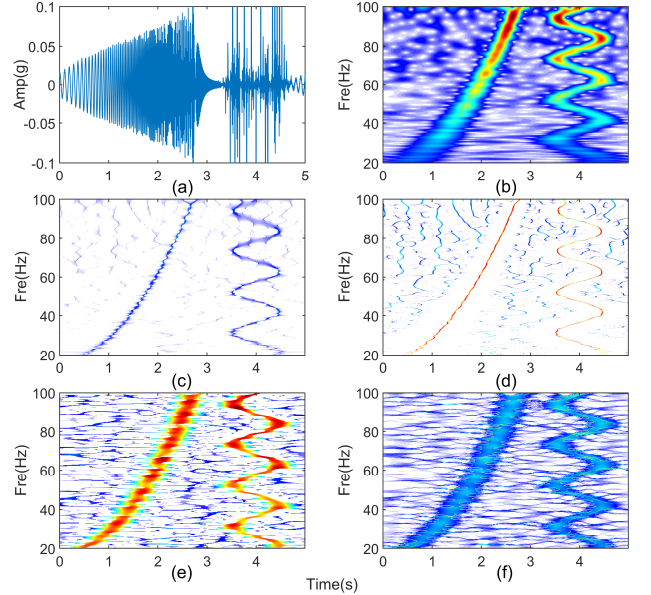

**Fig. 5.** (a) WT result, (b) WTSST result, (c) WTMSST result, (d) SET result and (e) SST result.

## V. REAL TRANSIENT SIGNAL ANALYSIS

In this section, we introduce several different real signals, i.e., bearing fault, ECG and GW signal. As an important part of rotating machinery, rolling bearing is a common cause of rotating machinery fault. Analyzing the bearing vibration signal is an effective way to describe and solve fault. ECG is one of the significant research objects in modern life science research. Through the statistics and analysis of ECG signal, the detailed theoretical basis of heart state can be obtained, which is beneficial to the objective diagnosis of heart diseases. And the exploration of GW signal enables us to reveal more unknow information in the universe. Therefore, we analyze a constant speed bearing fault signal, ECG data of patient with arrhythmia and a GW event to evaluate the performance of our proposed method.

### A. The Constant Speed Bearing Fault

The fault vibration signals of the bearing rotating at constant speed are collected from the rotating machinery test rig shown in Fig.6. It consists of an alternating current induction motor, a support shaft, a motor speed controller and a loading system. A radial load of 12 kN is provide to the rolling bearing LDK UER204 to speed up the test while ensuring that the shaft rotates at a constant frequency of 35 Hz [49]. In Fig.7, the vibration signal from operation to failure and the corresponding root mean square (RMS) value are recorded at a sampling frequency of 25600 Hz. It can be seen that after the experiment lasted for 80 minutes, the RMS value has a significant increase trend, which means that the fault is in the early stage. After 123 minutes, the bearing outer race is completely malfunctioning. Combined with the bearing parameters, theoretically, the fault characteristic frequency (FCF) at the current condition is 107.9 Hz, which means that the time interval between two continuous pulses is 9.3 ms.

The time-domain waveform fragment of the outer race fault signal in the early stage is depicted in Fig. 8(a). It can be seen that the fault signal is composed of a series of repetitive transient characteristics. WT result in Fig. 8(b) presents a blurred TFR, which only provides us a rough understanding of the signal characteristic. Fig. 8(e) and 8(f) prove that SET and SST are unable to appropriately describe the TFR of transient signal. The result of WTSST in Fig.8(c) shows that WTSST can provide an improved TFR, but it is still unable to accurately determine the bearing fault features. However, the WTMSST result in Fig.8(d) generates a highly concentrated TFR for each pulse component of the signal. Furthermore, the Rényi entropy values in Table I also show that the locating capacity of WTMSST is the most suitable choice among these approaches.

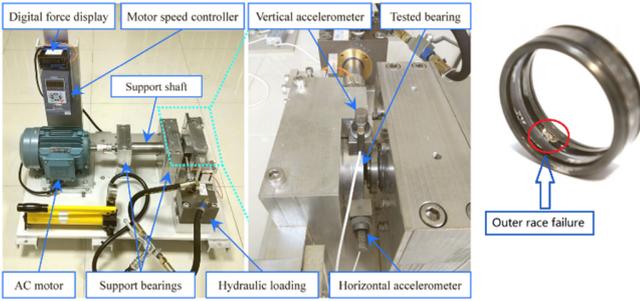

**Fig. 6.** The structural sketch of the experiment rig.

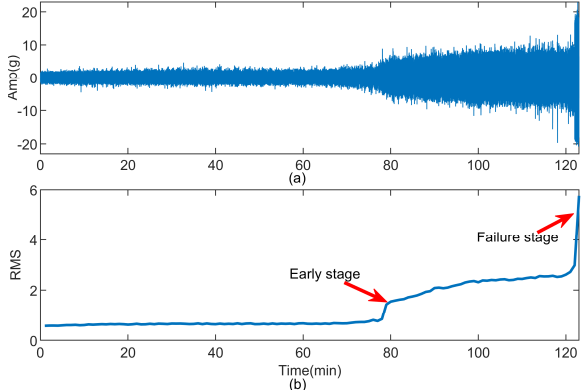

**Fig. 7.** (a) vibration signal with fault and (b) RMS.

In practical industrial application, the execution speed of diagnostic tool is also a significant index for real-time observation of mechanical faults. The measurement time of such algorithms is listed in Table II. The results illustrate that the complex postprocessing operation still enables the proposed method to execute in less than 1s as the other methods. Therefore, it can be concluded that the proposed method can provide the ability of real-time fault detection.

Generating a centralized TFR alone cannot meet the diagnostic requirements for fault signals with a lot of interference. Therefore, we need to extract the valuable pulse features from the signal to describe the bearing fault. The bearing fault signal of rotating machinery is usually composed of a series of periodic pulse components whose period depends on the bearing and fault type. Because the pulse signal usually occupies a large bandwidth, it should have a frequency point with prominent TF amplitude to characterize the pulse interval of the signal. The maximum of the envelope spectrogram of each frequency point in WTMSST is calculated as follows

$$TFES(\omega) = \max \left| \int_{-\infty}^{+\infty} S^{[N]}(\omega_0/\omega, b) - \phi(\omega) \right| e^{-i\xi b} db. \quad (36)$$

where $\phi(\omega)$ denotes the mean value of the WTMSST result in specified frequency point. Equation (36) is called TF envelop spectrum (TFES). Therefore, WTMSST considers the results at the frequency bin corresponding to the maximum TFES value to describe the fault characteristics. The TFES in Fig. 9(a) denotes that the TFR of WTMSST at 1060 Hz has obvious pulse feature. The slice representation at 1060 Hz is plotted in Fig. 9. The time interval of 9.3 ms between two continuous pulses is the same as the theoretical calculation, which means that the fault begins to occur. It is certain that the proposed method can still extract accurate pulse features to judge the existence of faults when the early fault information is weak. Therefore, the fault can be found effectively in advance by observing the characteristic frequency of early failure.

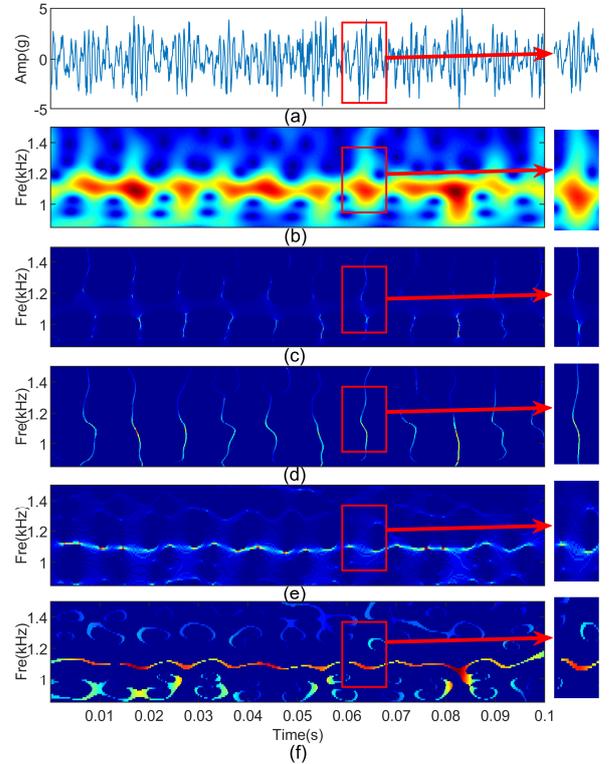

**Fig. 8.** (a) The fault signal waveform, (b) WT result, (c) WTSST result, (d) WTMSST result , (e) SST result and (f) SET result.

TABLE I
RÉNYI ENTROPIES OF THE SEVERAL METHODS

| TFA | WT | WTMSST(N=10) | STFT | SST | SET |
|---|---|---|---|---|---|
| Rényi Entropy | 18.37 | 11.59 | 18.01 | 15.42 | 13.50 |

TABLE II
REQUIRED EXECUTION TIME BY SEVERAL METHODS

| TFA | WT | WTSST | WTMSST(N=10) | SST | SET |
|---|---|---|---|---|---|
| Time (s) | 0.1094 | 0.2500 | 0.7656 | 0.2813 | 0.2969 |

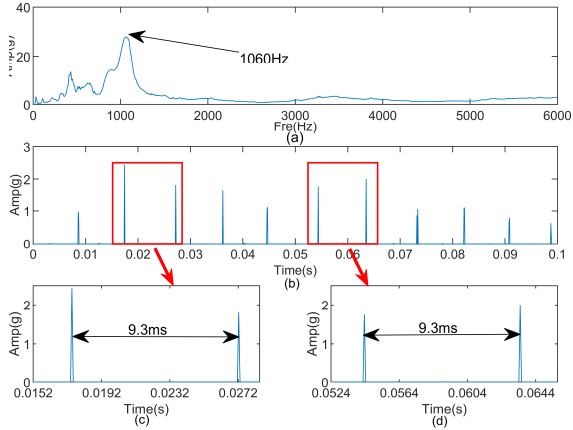

**Fig. 9.** (a) TFES, (b) the extracted impulse features and (c-d) the zoomed versions.

### B. Application to ECG Signal

In this subsection, we investigate the performance of our proposed techniques in coping with transient ECG signal. The ECG time series of patients with arrhythmia acquired by MIT-BIH [50] is shown in Fig. 10(a). And the TFA results in Fig. (10) illustrate that only WTMSST can interpret the transient feature of the ECG signal accurately, which allows us to observe the state of the heart more easily.

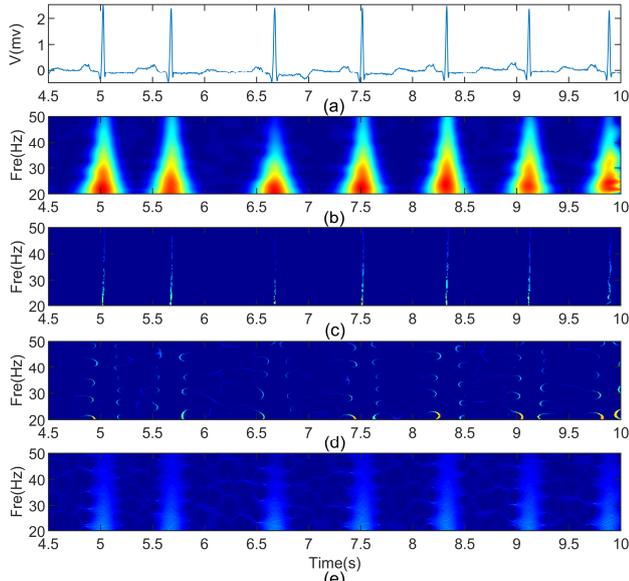

**Fig. 10.** (a) ECG waveform, (b) WT result, (c) WTMSST result, (d) SET result and (e) SST result.

As a disease type with periodic fluctuations in ECG, arrhythmia can be diagnosed by analyzing the interval between the positions where the TF energy of WT is most concentrated (i.e., the sharpened TFR of WTMSST). Fig. 11(c) illustrates that the average heartbeat cycle of the patient is 0.8708 s and there are two abnormal heartbeat points within 4-8 s, which shows that the patient has arrhythmia symptoms. Furthermore, the reconstruction performance of the WTMSST is also given in Fig. 11(a) and 11(b), which denotes the interest of the proposed approaches in real applications. Therefore, we can conclude that the proposed method is useful to extract the transient features in ECG signal.

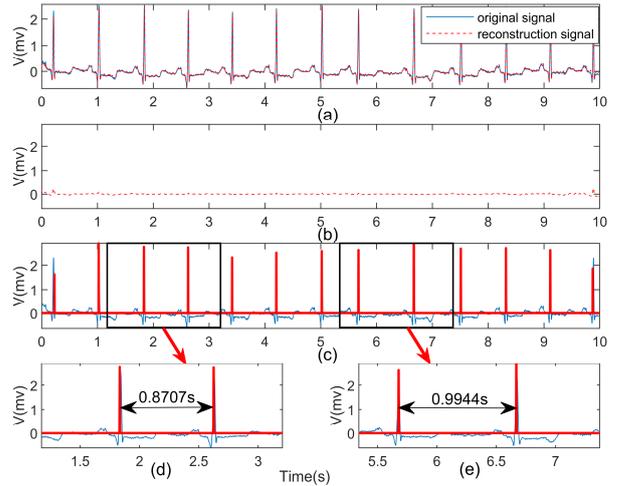

**Fig. 11.** (a) The reconstruction comparison, (b) the reconstruction error, (c) the extracted impulse features and (d-e) the zoomed versions.

### C. Application to GW Event

The GW event named GW190521 is introduced in this subsection to evaluate the applicability of our proposed method. This event detected by the LIGO detector Livingston is a transient GW signal generated by the merger of two intermediate mass black holes [3]. As a short transient signal, the duration of GW190521 event lasted only 0.1s. The whitened signal and GW waveform model and the TFRs of WT, WTSST and WTMSST are shown in Fig.12. The strong assumption framework of WTSST limits its application in GW signal. The sharpened TFR of WTMSST clearly illustrates the transient feature of GW signal. Fig.13 displays the reconstruction time series from WTMSST and the residual error. The concentrated TFR and the reasonable accuracy denote the effectiveness of our proposed method in GW signal detection.

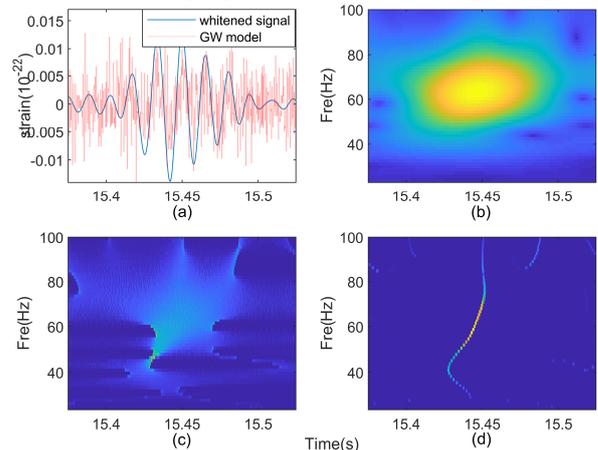

**Fig. 12.** (a) GW waveform, (b) WT result (c) WTSST result and (d) WTMSST result.

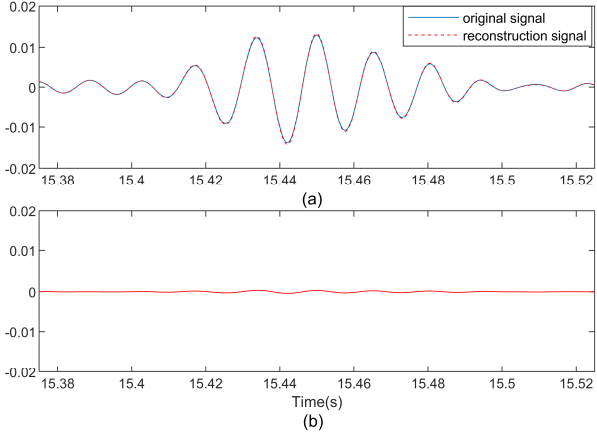

**Fig. 13.** (a) The reconstruction comparison and (b) error.

## VI. CONCLUSION

A postprocessing of WT method with high resolution is proposed in this paper. By squeezing the WT coefficients along the time direction, the WT is more suitable for processing the transient signal with strong amplitude modulation. Moreover, the iterative acquisition of GD candidate navigates the squeezing operation to a more precise location, which effectively enhances the energy concentration of WT. Meanwhile, we also prove that the proposed method still allows signal reconstruction. And we provide the detailed theoretical analysis of the proposed method to support its application. The experimental data of synthetic numerical signal and different real signal are used to evaluate the performance of the proposed method. The results show that compared with other advanced TFA tools, the proposed method has more concentrated TFR when dealing with signals rich in transient events.

## APPENDIX A

### PROOF OF REASSIGNMENT CANDIDATES

In this Appendix, we will give the specific derivation of $\hat{\omega}(a,b)$ and $\hat{t}(a,b)$ under MWT using the probability distribution function $Q_{a,b}(t,\xi)$ around the point $(a,b)$. The $\hat{\omega}(a,b)$ and $\hat{t}(a,b)$, which are centers of gravity of energy, can be defined by the following weighted integral

$$\hat{\omega}(a,b) = \frac{\iint \xi Q_{a,b}(t,\xi) dt d\xi}{\iint Q_{a,b}(t,\xi) dt d\xi}, \quad \hat{t}(a,b) = \frac{\iint t Q_{a,b}(t,\xi) dt d\xi}{\iint Q_{a,b}(t,\xi) dt d\xi}. \tag{A1}$$

And then we have

$$\hat{\omega}(a,b) = \frac{\iint \xi Q_{a,b}(t,\xi) dt d\xi}{\iint Q_{a,b}(t,\xi) dt d\xi} = \frac{W_e(a,b)\iint \xi \hat{x}^*(\xi)\hat{g}^*\left(a(\xi-\omega)\right)e^{-i(\xi-\omega)b} dt d\xi}{2\pi W_e(a,b)\overline{W_e(a,b)}}$$

$$= \omega + \frac{\iint i(\xi-\omega)\hat{x}(\xi)\hat{g}\left(a(\xi-\omega)\right)e^{i(\xi-\omega)b} dt d\xi}{-i2\pi \overline{W_e(a,b)}} = \omega + \frac{\partial_b \overline{W_e(a,b)}}{i\overline{W_e(a,b)}} \tag{A2}$$

Since the IF candidate is the real part of the center of gravity of energy, it yields

$$\hat{\omega}(a,b) = \Re\left[\omega + \frac{\partial_b \overline{W_e(a,b)}}{i\overline{W_e(a,b)}}\right] = \Re\left[\omega + \frac{\partial_b W_e(a,b)}{i W_e(a,b)}\right] \tag{A3}$$

Similarly, the GD candidate can be obtained as

$$\hat{t}(a,b) = \frac{\iint t Q_{a,b}(t,\xi) dt d\xi}{\iint Q_{a,b}(t,\xi) dt d\xi} = \frac{\iint t x(t) g\left((t-b)/a\right) e^{-i\omega t} / a \, dt d\xi}{W_e(a,b)} \tag{A4}$$

## APPENDIX B

### PROOF OF THEOREM 1

The realization of the main results in Theorem 1 depends on a series of estimates, which we will give and prove under the framework assumed in Definition 1.

**Estimate 1.** For each component i.e., $k \in \{1,\ldots,K\}$, we can get

$$|A_k(\xi+\omega) - A_k(\omega)| \leq \varepsilon\left(|\varphi_k'(\omega)||\xi| + \frac{1}{2}M''|\xi|^2\right) \text{ and } |\varphi_k'(\xi+\omega) - \varphi_k'(\omega)| \leq \varepsilon\left(|\varphi_k'(\omega)||\xi| + \frac{1}{2}M''|\xi|^2\right).$$

**Proof.** For any $\omega \geq 0$, we have

$$\left|A_k(\xi+\omega)-A_k(\omega)\right|=\left|\int_0^\xi A_k'(\omega+u)du\right|\le\int_0^\xi\left|A_k'(\omega+u)\right|du\le\varepsilon\int_0^\xi\left|\varphi_k'(\omega+u)\right|du=\varepsilon\int_0^\xi\left|\varphi_k'(\omega)+\int_0^u\varphi_k''(\omega+x)dx\right|du$$

$$\le\varepsilon\left(\left|\varphi_k'(\omega)\right|\left|\xi\right|+\frac{1}{2}M''\left|\xi\right|^2\right)$$

Similarly, we can prove another inequality with the above proof. Next, we will prove that when $|b+\varphi_k'(\omega)|>\Delta$, $|W_{e,x}^{\hat{g}}(a,b)|\le\varepsilon E_0(\omega)$ holds for all $k\in\{1,\ldots,K\}$. For convenience, let

$$Q_{k,1}(a,b)=\sum_{k=1}^K\frac{1}{2\pi}\int_{\mathbb{R}^+}A_k(\omega)e^{i(\varphi_k(\omega)+\varphi_k'(\omega)(\xi-\omega))}\hat{g}(a(\xi-\omega))e^{i(\xi-\omega)b}d\xi$$
$$=\sum_{k=1}^K A_k(\omega)e^{i\varphi_k(\omega)}g\left((b+\varphi_k'(\omega))/a\right)/a.$$
(B1)

**Estimate 2.** For $k\in\{1,\ldots,K\}$ and $(a,b)\in\mathbb{R}^+\times\mathbb{R}$, we have
$$\left|W_{e,f}^{\hat{g}}(a,b)-Q_{k,1}(a,b)\right|\le\varepsilon E_0(a)$$
where
$$E_0(a)=\frac{1}{2\pi}\sum_{k=1}^K I_1\left|a^{-2}\varphi_k'(\omega)\right|+\sum_{k=1}^K\frac{1}{2}I_2\left|a^{-3}\right|\left[\left|A_k(\omega)\right|\left|\varphi_k'(\omega)\right|+M''\right]+\sum_{k=1}^K\frac{1}{6}I_3\left|a^{-4}A_k(\omega)\right|M''\text{ with }I_n=\int|x|^n\left|\hat{g}(x)\right|dx.$$

**Proof.** According to $\hat{x}(\omega)\in A_{\varepsilon,d}$ and $\hat{g}\in S$, the MWT of $\hat{x}(\omega)=\sum_{k=1}^K A_k(\omega)e^{i\varphi_k(\omega)}$ can be well defined as

$$W_{e,x}^{\hat{g}}(a,b)=\frac{1}{2\pi}\sum_{k=1}^K\int_{\mathbb{R}^+}A_k(\xi)e^{i\varphi_k(\xi)}\hat{g}(a(\xi-\omega))e^{i(\xi-\omega)b}d\xi$$
$$=\frac{1}{2\pi}\sum_{k=1}^K\int_{\mathbb{R}^+}A_k(\omega)e^{i\left(\varphi_k(\omega)+\varphi_k'(\omega)(\xi-\omega)+\int_0^{\xi-\omega}\varphi_k'(\omega+u)-\varphi_k'(\omega)du\right)}\hat{g}(a(\xi-\omega))e^{i(\xi-\omega)b}d\xi$$
$$+\frac{1}{2\pi}\sum_{k=1}^K\int_{\mathbb{R}^+}\left[A_k(\xi)-A_k(\omega)\right]e^{i\varphi_k(\xi)}\hat{g}(a(\xi-\omega))e^{i(\xi-\omega)b}d\xi$$

Appling $Q_{k,1}(a,b)$, we have

$$\left|W_{e,x}^{\hat{g}}(a,b)-Q_{k,1}(a,b)\right|\le\frac{1}{2\pi}\sum_{k=1}^K\int_{\mathbb{R}^+}\left|A_k(\xi)-A_k(\omega)\right|\left|\hat{g}(a(\xi-\omega))\right|d\xi$$
$$+\frac{1}{2\pi}\sum_{k=1}^K\int_{\mathbb{R}^+}\left|A_k(\omega)\right|\left|e^{i\left(\varphi_k(\omega)+\varphi_k'(\omega)(\xi-\omega)+\int_0^{\xi-\omega}\varphi_k'(\omega+u)-\varphi_k'(\omega)du\right)}-e^{i(\varphi_k(\omega)+\varphi_k'(\omega)(\xi-\omega))}\right|\left|\hat{g}(a(\xi-\omega))\right|d\xi$$
$$\le\frac{1}{2\pi}\sum_{k=1}^K\varepsilon\int_{\mathbb{R}^+}\left|\xi-\omega\right|\left(\left|\varphi_k'(\omega)\right|+\frac{1}{2}M_k''\left|\xi-\omega\right|\right)\left|\hat{g}(a(\xi-\omega))\right|d\xi$$
$$+\frac{1}{2\pi}\sum_{k=1}^K\left|A_k(\omega)\right|\int_{\mathbb{R}^+}\left|\int_0^{\xi-\omega}\varphi_k'(\omega+u)-\varphi_k'(\omega)du\right|\left|\hat{g}(a(\xi-\omega))\right|d\xi$$
$$\le\frac{1}{2\pi}\sum_{k=1}^K\varepsilon\int_{\mathbb{R}^+}\left|\xi-\omega\right|\left(\left|\varphi_k'(\omega)\right|+\frac{1}{2}M_k''\left|\xi-\omega\right|\right)\left|\hat{g}(a(\xi-\omega))\right|d\xi$$
$$+\frac{1}{2\pi}\sum_{k=1}^K\varepsilon\left|A_k(\omega)\right|\int_{\mathbb{R}^+}\left[\frac{1}{2}\left|\xi-\omega\right|^2\left|\varphi_k'(\omega)\right|+\frac{1}{6}\left|\xi-\omega\right|^3 M_k''\right]\left|\hat{g}(a(\xi-\omega))\right|d\xi$$
$$=\frac{\varepsilon}{2\pi}\left(\sum_{k=1}^K I_1\left|a^{-2}\varphi_k'(\omega)\right|+\sum_{k=1}^K\frac{1}{2}I_2\left|a^{-3}\right|\left[\left|A_k(\omega)\right|\left|\varphi_k'(\omega)\right|+M''\right]+\sum_{k=1}^K\frac{1}{6}I_3\left|a^{-4}A_k(\omega)\right|M''\right)$$

i.e., $\left|W_{e,x}^{\hat{g}}(a,b)-Q_{k,1}(a,b)\right|\le\varepsilon E_0(a)$.

**Lemma 1.** For any $(a,b)$ in TS plane, there is most one $k\in\{1,\ldots,K\}$ satisfying $|b+\varphi_k'(\omega)|<\Delta$.

**Proof.** Suppose that there exists $k,l\in\{1,\ldots,K\}$ satisfying $|b+\varphi_k'(\omega)|<\Delta$, i.e., when $k\ne l$, $|b+\varphi_k'(\omega)|<\Delta$ and $|b+\varphi_l'(\omega)|<\Delta$. Without losing generality, let $k>l$, then, we have
$$\varphi_k'(\omega)-\varphi_l'(\omega)\ge\varphi_k'(\omega)-\varphi_{k-1}'(\omega)\ge 2\Delta.$$
According compactly supported domain, we have
$$-\Delta-b<\varphi_k'(\omega)<\Delta-b,\quad-\Delta-b<\varphi_l'(\omega)<\Delta-b,$$

which yields
$\varphi_k'(\omega) - \varphi_l'(\omega) < 2\Delta$.

It is obvious that this result is contrary to the well separated condition. Therefore, only one $k \in \{1,\ldots,K\}$, i.e., $k = l$, allows $|b + \varphi_k'(\omega)| < \Delta$. According to the Estimate 2 and Lemma 1, the phenomenon that the TS plane is separated into K non-contact subregions by $Z_k := \{|b + \varphi_k'(\omega)| < \Delta\}$ and the $|W_{e,x}^{\hat{g}}(a,b)|$ between the adjacent subregions is sufficiently small is revealed. If an additional restriction is imposed on $\varepsilon$: for all $k \in \{1,\ldots,K\}$ and all pairs $(a,b) \in Z_k$

$$\varepsilon \leq E_0^{-2/3}(a), \tag{B2}$$

then $\varepsilon E_0(a) \leq \tilde{\varepsilon}$ will be obtained. Therefore, for any $k \in \{1,\ldots,K\}$ and any $(a,b) \in Z_k$, we have $\left|W_{e,x}^{\hat{g}}(a,b) - Q_{k,1}(a,b)\right| \leq \varepsilon E_0(a) \leq \tilde{\varepsilon}$ and for any $(a,b) \notin Z_k$, we have $|W_{e,x}^{\hat{g}}(a,b)| \leq \varepsilon E_0(a) \leq \tilde{\varepsilon}$. According to the expression of $\hat{t}(a,b)$, it is necessary to estimate $G_x^{\hat{g}}(a,b) = \sum_{l=1}^{K} \int_{\mathbb{R}^+} x_l'(\xi) \hat{g}(a(\xi-\omega)) e^{i(\xi-\omega)b} d\xi$ in the specific zones.

**Estimate 3.** For $k \in \{1,\ldots,K\}$ and $(a,b) \in \mathbb{R}^+ \times \mathbb{R}$ such that $|b + \varphi_k'(\omega)| < \Delta$, we have

$$\left|iG_x^g(a,b) + Q_{k,1}(a,b)(b + \varphi'(\omega) - b)\right| \leq \varepsilon \Gamma_0(a,b)$$

where

$$\Gamma_0(a,b) = \frac{1}{2\pi} \sum_{k=1}^{K} I_1' |a^{-1}| |\varphi_k'(\omega)| + \sum_{k=1}^{K} \frac{1}{2} I_2' |a^{-2}| \left[|A_k(\omega)||\varphi_k'(\omega)| + M''\right] + \sum_{k=1}^{K} \frac{1}{6} I_3' |a^{-3}| M'' |A_k(\omega)| + bE_0(a) \text{ with } I_n' = \int |x|^n g'(x) dx.$$

**Proof.** We first give the concrete representation of $G_x^{\hat{g}}(a,b)$, i.e.,

$$iG_x^{\hat{g}}(a,b) = \sum_{l=1}^{K} \int_{\mathbb{R}^+} x_l'(\xi) \hat{g}(a(\xi-\omega)) e^{i(\xi-\omega)b} d\xi$$

$$= -i\sum_{l=1}^{K} \int_{\mathbb{R}^+} A_l(\xi) e^{i\varphi_l(\xi)} a\hat{g}'(a(\xi-\omega)) e^{i(\xi-\omega)b} d\xi + b\sum_{l=1}^{K} \int_{\mathbb{R}^+} A_l(\xi) e^{i\varphi_l(\xi)} \hat{g}(a(\xi-\omega)) e^{i(\xi-\omega)b} d\xi$$

$$= -i\sum_{l=1}^{K} \int_{\mathbb{R}^+} A_l(\omega) e^{i\left(\varphi_l(\omega) + \varphi_l'(\omega)(\xi-\omega) + \int_0^{\xi-\omega} \varphi_l'(\omega+u) - \varphi_l'(\omega) du\right)} a\hat{g}(a(\xi-\omega)) e^{i(\xi-\omega)b} d\xi$$

$$-i\sum_{l=1}^{K} \int_{\mathbb{R}^+} \left[A_l(\xi) - A_l(\omega)\right] e^{i\varphi_l'(\xi)} a\hat{g}(a(\xi-\omega)) e^{i(\xi-\omega)b} d\xi + b\sum_{l=1}^{K} \int_{\mathbb{R}^+} A_l(\xi) e^{i\varphi_l(\xi)} \hat{g}(a(\xi-\omega)) e^{i(\xi-\omega)b} d\xi.$$

According to the lemma 1, there is only one term k i.e., $k = l$ satisfies $|b + \varphi_k'(\omega)| < \Delta$. Therefore, we can obtain

$$iG_x^{\hat{g}}(a,b) = -i\sum_{k=1}^{K} \int_{\mathbb{R}^+} A_k(\omega) e^{i\left(\varphi_k(\omega) + \varphi_k'(\omega)(\xi-\omega) + \int_0^{\xi-\omega} \varphi_k'(\omega+u) - \varphi_k'(\omega) du\right)} a\hat{g}'(a(\xi-\omega)) e^{i(\xi-\omega)b} d\xi$$

$$-i\sum_{k=1}^{K} \int_{\mathbb{R}^+} \left[A_k(\xi) - A_k(\omega)\right] e^{i\varphi_k'(\xi)} a\hat{g}'(a(\xi-\omega)) e^{i(\xi-\omega)b} d\xi + b\sum_{k=1}^{K} \int_{\mathbb{R}^+} A_k(\xi) e^{i\varphi_k(\xi)} \hat{g}(a(\xi-\omega)) e^{i(\xi-\omega)b} d\xi$$

and we also note that

$$Q_{k,1}(a,b)(b + \varphi'(\omega) - b) = \sum_{k=1}^{K} \int_{\mathbb{R}^+} (b + \varphi'(\omega)) A_k(\omega) e^{i(\varphi_k(\omega) + \varphi_k'(\omega)(\xi-\omega))} \hat{g}(a(\xi-\omega)) e^{i(\xi-\omega)b} d\xi$$

$$- \sum_{k=1}^{K} b\int_{\mathbb{R}^+} A_k(\omega) e^{i(\varphi_k(\omega) + \varphi_k'(\omega)(\xi-\omega))} \hat{g}(a(\xi-\omega)) e^{i(\xi-\omega)b} d\xi \int_{\mathbb{R}^+}$$

$$= \sum_{k=1}^{K} A_k(\omega) e^{i\varphi_k(\omega)} \int_{\mathbb{R}^+} (b + \varphi'(\omega)) \hat{g}(a(\xi-\omega)) e^{i(b+\varphi_k'(\omega))(\xi-\omega)} d\xi$$

$$- \sum_{k=1}^{K} b\int_{\mathbb{R}^+} A_k(\omega) e^{i(\varphi_k(\omega) + \varphi_k'(\omega)(\xi-\omega))} \hat{g}(\xi-\omega) e^{i(\xi-\omega)b} d\xi$$

$$= \sum_{k=1}^{K} iA_k(\omega) e^{i\varphi_k(\omega)} \int_{\mathbb{R}^+} a\hat{g}'(a(\xi-\omega)) e^{i(b+\varphi_k'(\omega))(\xi-\omega)} d\xi$$

$$- \sum_{k=1}^{K} b\int_{\mathbb{R}^+} A_k(\omega) e^{i(\varphi_k(\omega) + \varphi_k'(\omega)(\xi-\omega))} \hat{g}(a(\xi-\omega)) e^{i(\xi-\omega)b} d\xi$$

The above two parts together constitute the proof of Estimate 3. And we can give the following inequality

$$\left| iG_f^{\hat{g}}(a,b) + Q_{k,1}(a,b)(b + \varphi'(\omega) - b) \right|$$

$$= \left| \frac{1}{2\pi} \sum_{k=1}^{K} i \int_{\mathbb{R}^+} A_k(\omega) \left( e^{i\left(\varphi_k(\omega) + \varphi'_k(\omega)(\xi - \omega) + \int_0^{\xi-\omega} \varphi'_k(\omega+u) - \varphi'_k(\omega) du\right)} - e^{i\varphi_k(\omega) + \varphi'_k(\omega)(\xi - \omega)} \right) a\hat{g}'(a(\xi - \omega)) e^{i(\xi-\omega)b} d\xi \right.$$
$$\left. + \frac{1}{2\pi} \sum_{k=1}^{K} i \int_{\mathbb{R}^+} \left[ A_k(\xi) - A_k(\omega) \right] e^{i\varphi'_k(\xi)} a\hat{g}'(a(\xi-\omega)) e^{i(\xi-\omega)b} d\xi + b\left( W_{e,x}^{g}(a,b) - Q_{k,1}(a,b) \right) \right|$$

$$\leq \frac{1}{2\pi} \sum_{k=1}^{K} |A_k(\omega)| \int_{\mathbb{R}^+} \left| \int_0^{\xi-\omega} \varphi'_k(\omega+u) - \varphi'_k(\omega) du \right| |a\hat{g}'(a(\xi-\omega))| d\xi$$
$$+ \frac{1}{2\pi} \sum_{k=1}^{K} \int_{\mathbb{R}^+} |A_k(\xi) - A_k(\omega)| |a\hat{g}'(a(\xi-\omega))| d\xi + b\varepsilon E_0(a)$$

$$\leq \frac{1}{2\pi} \sum_{k=1}^{K} \varepsilon |A_k(\omega)| \int_{\mathbb{R}^+} \left[ \frac{1}{2} |\xi-\omega|^2 |\varphi'_k(\omega)| + \frac{1}{6} |\xi-\omega|^3 M'' \right] |a\hat{g}'(a(\xi-\omega))| d\xi$$
$$+ \frac{1}{2\pi} \sum_{k=1}^{K} \varepsilon \int_{\mathbb{R}^+} |\xi-\omega| \left( |\varphi'_k(\omega)| + \frac{1}{2} M'' |\xi-\omega| \right) |a\hat{g}'(a(\xi-\omega))| d\xi + b\varepsilon E_0(a)$$

$$= \frac{\varepsilon}{2\pi} \left( \sum_{k=1}^{K} I'_1 |a^{-1}| |\varphi'_k(\omega)| + \sum_{k=1}^{K} \frac{1}{2} I'_2 |a^{-2}| \left[ |A_k(\omega)| |\varphi'_k(\omega)| + M'' \right] + \sum_{k=1}^{K} \frac{1}{6} I'_3 |a^{-3}| M'' |A_k(\omega)| + bE_0(a) \right)$$

**Estimate 4.** For any $k \in \{1,\ldots,K\}$ and any $(a,b) \in Z_k$, such that $|W_{e,x}^{\hat{g}}(a,b)| \geq \tilde{\varepsilon}$, we have

$$\left| \hat{t}(a,b) + \varphi'_k(\omega) \right| \leq \varepsilon^{2/3} \left( \Gamma_0(a,b) + E_0(a) \varphi'_k(\omega) \right).$$

**Proof.** According to the definition of $\hat{t}(a,b)$

$$\hat{t}(a,b) = \frac{i \int_{\mathbb{R}^+} \hat{x}'(\xi) g(a(\xi-\omega)) e^{i(\xi-\omega)b} d\xi}{\int_{\mathbb{R}^+} \hat{x}(\xi) g(a(\xi-\omega)) e^{i(\xi-\omega)b} d\xi},$$

we can obtain

$$\hat{t}(a,b) + \varphi'(\omega) = \frac{i \int_{\mathbb{R}^+} \hat{x}'(\xi) g(a(\xi-\omega)) e^{i(\xi-\omega)b} d\xi}{\int_{\mathbb{R}^+} \hat{x}(\xi) g(a(\xi-\omega)) e^{i(\xi-\omega)b} d\xi} + \varphi'(\omega)$$

$$= \frac{-i \int_{\mathbb{R}^+} \hat{x}(\xi) ag'(a(\xi-\omega)) e^{i(\xi-\omega)b} d\xi + \int_{\mathbb{R}^+} \hat{x}(\xi) b g(a(\xi-\omega)) e^{i(\xi-\omega)b} d\xi}{\int_{\mathbb{R}^+} \hat{x}(\xi) g(a(\xi-\omega)) e^{i(\xi-\omega)b} d\xi} + \frac{\varphi'(\omega) \int_{\mathbb{R}^+} \hat{x}(\xi) g(a(\xi-\omega)) e^{i(\xi-\omega)b} d\xi}{\int_{\mathbb{R}^+} \hat{x}(\xi) g(a(\xi-\omega)) e^{i(\xi-\omega)b} d\xi}$$

Considering the Estimate 2 and the Estimate 3 yields

$$\left| \hat{t}(a,b) + \varphi'_k(\omega) \right| \leq \left| \frac{iG_{e,x}^{\hat{g}}(a,b) + (b + \varphi'_k(\omega) - b) Q_{k,1}(a,b)}{W_{e,x}^{\hat{g}}(a,b)} \right| + \left| \frac{\left( W_{e,x}^{\hat{g}}(a,b) - Q_{k,1}(a,b) \right) \varphi'_k(\omega)}{W_{e,x}^{\hat{g}}(a,b)} \right|$$

$$\leq \frac{\varepsilon \Gamma_0(a,b) + \varepsilon E_0(a) \varphi'_k(\omega)}{|W_{e,x}^{\hat{g}}(a,b)|} \leq \varepsilon^{2/3} \left( \Gamma_0(a,b) + E_0(a) \varphi'_k(\omega) \right)$$

For all $k \in \{1,\ldots,K\}$ and all $(a,b) \in \{ | W_{e,x}^{\hat{g}}(a,b) | \geq \tilde{\varepsilon}; |b + \varphi'(\omega) | < \Delta \} \subset Z_k$, the additional restriction on $\varepsilon$ is considered, i.e.,

$$0 < \varepsilon < \left( \Gamma_0(a,b) + E_0(a) \varphi'_k(\omega) \right)^{-3} \tag{B3}$$

then, the error result is given: $\left| \hat{t}(a,b) + \varphi'_k(\omega) \right| \leq \tilde{\varepsilon}$.

So far, the all items in Theorem 1 have given a detailed theoretical analysis, which ensures that WTSST can estimate accurately when processing weak frequency-varying signal.

## APPENDIX C

### PROOF OF GD ACCURACY

If we specify the window as Gaussian function in (20), i.e., $\hat{g}(\omega) = \sqrt{2\pi}\sigma e^{i0.5\sigma\omega^2}$ and the signal can be expressed as $\hat{x}(\xi) = A(\omega) e^{i(\varphi(\omega) + \varphi'(\omega)(\xi-\omega) + 0.5\varphi''(\omega)(\xi-\omega)^2)}$, we can rewrite the (20) as

$$\hat{t}(a,b) = -\varphi'(\omega) - \varphi''(\omega)\frac{W^{\xi\hat{g}}}{aW^{\hat{g}}}$$

$$= -\varphi'(\omega) - \varphi''(\omega)\frac{\int_0^{+\infty}(\xi-\omega)A(\omega)e^{i(\varphi(\omega)+\varphi'(\omega)(\xi-\omega)+0.5\varphi''(\omega)(\xi-\omega)^2)}e^{i\sigma a^2(\xi-\omega)^2}e^{i(\xi-\omega)b}d\xi}{\int_0^{+\infty}A(\omega)e^{i(\varphi(\omega)+\varphi'(\omega)(\xi-\omega)+0.5\varphi''(\omega)(\xi-\omega)^2)}e^{i\sigma a^2(\xi-\omega)^2}e^{i(\xi-\omega)b}d\xi} \quad (C1)$$

$$= -\varphi'(\omega) - \varphi''(\omega)\frac{\int_0^{+\infty} xe^{i(\varphi'(\omega)+b)x+0.5(i\varphi''(\omega)-a^2\sigma)x^2}d\xi}{\int_0^{+\infty} e^{i(\varphi'(\omega)+b)x+0.5(i\varphi''(\omega)-a^2\sigma)x^2}d\xi}$$

For the numerator of C1, we have

$$\int_0^{+\infty} xe^{i(\varphi'(\omega)+b)x+0.5(i\varphi''(\omega)-a^2\sigma)x^2}d\xi = \int_0^{+\infty}\left(x+\frac{i(\varphi'(\omega)+b)}{i\varphi''(\omega)-a^2\sigma}-\frac{i(\varphi'(\omega)+b)}{i\varphi''(\omega)-a^2\sigma}\right)e^{(0.5(i\varphi''(\omega)-a^2\sigma))\left[\left(x+\frac{i(\varphi'(\omega)+b)}{i\varphi''(\omega)-a^2\sigma}\right)^2-\left(\frac{i(\varphi'(\omega)+b)}{i\varphi''(\omega)-a^2\sigma}\right)^2\right]}d\xi$$

$$= e^{\frac{(\varphi'(\omega)+b)^2}{i\varphi''(\omega)-a^2\sigma}}\int_0^{+\infty}\xi e^{(0.5(i\varphi''(\omega)-a^2\sigma))\xi^2}d\xi - \frac{i(\varphi'(\omega)+b)}{i\varphi''(\omega)-a^2\sigma}e^{\frac{(\varphi'(\omega)+b)^2}{i\varphi''(\omega)-a^2\sigma}}\int_0^{+\infty}e^{(0.5(i\varphi''(\omega)-a^2\sigma))\xi^2}d\xi$$

$$= \frac{i(\varphi'(\omega)+b)}{i\varphi''(\omega)-a^2\sigma}e^{\frac{(\varphi'(\omega)+b)^2}{i\varphi''(\omega)-a^2\sigma}}\frac{\sqrt{\sigma}}{\sqrt{a^2\sigma-i\varphi''(\omega)}}$$

Similarly, for the denominator of C1, we can obtain

$$\int_0^{+\infty} e^{i(\varphi'(\omega)+b)x+0.5(i\varphi''(\omega)-a^2\sigma)x^2}dx = \int_0^{+\infty} e^{(0.5(i\varphi''(\omega)-a^2\sigma))\left[\left(x+\frac{i(\varphi'(\omega)+b)}{i\varphi''(\omega)-a^2\sigma}\right)^2-\left(\frac{i(\varphi'(\omega)+b)}{i\varphi''(\omega)-a^2\sigma}\right)^2\right]}dx$$

$$= e^{\frac{(\varphi'(\omega)+b)^2}{i\varphi''(\omega)-a^2\sigma}}\int_0^{+\infty} e^{(0.5(i\varphi''(\omega)-a^2\sigma))\xi^2}d\xi$$

$$= -e^{\frac{(\varphi'(\omega)+b)^2}{i\varphi''(\omega)-a^2\sigma}}\frac{\sqrt{\sigma}}{\sqrt{a^2\sigma-i\varphi''(\omega)}}$$

Therefore, taking the real part of C1 yields

$$\hat{t}(a,b) = \Re\left[-\varphi'(\omega)+\varphi''(\omega)\frac{i(\varphi'(\omega)+b)}{i\varphi''(\omega)-a^2\sigma}\right] = -\varphi'(\omega) + \frac{\varphi''(\omega)^2}{\varphi''(\omega)^2+(a^2\sigma)^2}(\varphi'(\omega)+b) \quad (C2)$$

## APPENDIX D

### PROOF OF RECONSTRUCTION ACCURACY

This Appendix analyzes the reconstruction performance of WTSST and WTMSST. The related estimate and proof process are as follows:

**Estimate 5.** Suppose that (18) and (19) hold, and for all $a \in \mathbb{R}^+$, we assume that $\varepsilon \leq \Delta^3/8$. Then, for $\forall a \in \mathbb{R}^+$ and $\forall k \in \{1,\ldots,K\}$, we have

$$\left|\lim_{\delta\to 0}\frac{1}{\hat{g}(0)}\int_{|u+\varphi'(\omega)|<\tilde{\varepsilon}}S_{f,\tilde{\varepsilon}}^\delta(a,u)du - A_k(\omega)e^{i\varphi_k(\omega)}\right| \leq C\tilde{\varepsilon}.$$

**Proof.** For a fixed $a \in \mathbb{R}^+$, because $W_{e,x}^{\hat{g}}(a,b) \in C^\infty(A_{x,\tilde{\varepsilon}})$ (where $A_{x,\tilde{\varepsilon}}$ is a subset of compactly supported sets $\bigcup_{l=1}^K Z_l$), clearly $S_{x,\tilde{\varepsilon}}^\delta(a,u) \in C^\infty(\mathbb{R})$. Hence, we have

$$\lim_{\delta\to 0}\int_{|u+\varphi'(\omega)|<\tilde{\varepsilon}}S_{x,\tilde{\varepsilon}}^\delta(a,u)du = \lim_{\delta\to 0}\int_{|u+\varphi'(\omega)|<\tilde{\varepsilon}}\int_{A_{f,\tilde{\varepsilon}}(\omega)}W_{e,x}^{\hat{g}}(a,b)\frac{1}{\delta}\hat{h}\left(\frac{u-\hat{t}(a,b)}{\delta}\right)dbdu$$

$$= \int_{A_{f,\tilde{\varepsilon}}(\omega)}W_{e,x}^{\hat{g}}(a,b)\lim_{\delta\to 0}\int_{|u+\varphi'(\omega)|<\tilde{\varepsilon}}\frac{1}{\delta}\hat{h}\left(\frac{u-\hat{t}(a,b)}{\delta}\right)dbdu \quad (D1)$$

$$= \int_{A_{f,\tilde{\varepsilon}}(\omega)\cap\{b:|\hat{t}(a,b)+\varphi_k'(\omega)|<\tilde{\varepsilon}\}}W_{e,x}^{\hat{g}}(a,b)db$$

From Lemma 1, we can find that there is only one $l \in \{1,\ldots,K\}$ such that $|b+\varphi_k'(\omega)| < \Delta$, and since $A_{x,\tilde{\varepsilon}}$ is a subset of compactly supported sets $\bigcup_{l=1}^{K} Z_l$. If $l \neq k$, then there is

$$\left| \hat{t}(a,b) + \varphi_k'(\omega) \right| \geq \left| \varphi_k'(\omega) - \varphi_l'(\omega) \right| - \left| \hat{t}(a,b) + \varphi_l'(\omega) \right| \geq \tilde{\varepsilon}.$$

It means that there is only one $l \in \{1,\ldots,K\}$ satisfies $|b + \varphi_l'(\omega)| < \Delta$ and $|\hat{t}(a,b) + \varphi_k'(\omega)| < \tilde{\varepsilon}$ simultaneously, i.e., $l = k$. Therefore, the D1 can be write as

$$\lim_{\delta \to 0} \int_{|u+\varphi'(\omega)|<\tilde{\varepsilon}} S_{x,\tilde{\varepsilon}}^{\delta}(a,u) du = \int_{A_{f,\tilde{\varepsilon}}(\omega) \cap \{b:|b+\varphi_k'(\omega)|<\Delta\}} W_{e,x}^{\hat{g}}(a,b) db$$

$$= \int_{\{b:|b+\varphi_k'(\omega)|<\Delta\}} W_{e,x}^{\hat{g}}(a,b) db - \int_{\{b:|b+\varphi_k'(\omega)|<\Delta\}/A_{f,\tilde{\varepsilon}}(\omega)} W_{e,x}^{\hat{g}}(a,b) db$$

According to the B1, the $Q_{k,1}(a,b)$ can be expressed as follows

$$\int_{\{b:|b+\varphi_k'(\omega)|<\Delta\}} Q_{k,1}(a,b) db = \sum_{k=0}^{K} \int_{\{b:|b+\varphi_k'(\omega)|<\Delta\}} \frac{1}{2\pi} \int_{-\infty}^{+\infty} A_k(\omega) e^{i(\varphi_k(\omega)+\varphi_k'(\omega)(\xi-\omega))} \hat{g}(a(\xi-\omega)) e^{i(\xi-\omega)b} d\xi db$$

$$= \sum_{k=0}^{K} \frac{1}{2\pi} \int_{-\infty}^{+\infty} A_k(\omega) e^{i(\varphi_k(\omega)+\varphi_k'(\omega)(\xi-\omega))} \hat{g}(a(\xi-\omega)) \int_{\{b:|b+\varphi_k'(\omega)|<\Delta\}} e^{i(\xi-\omega)b} db d\xi$$

$$= \sum_{k=0}^{K} \hat{g}(0) A_k(\omega) e^{i\varphi_k(\omega)}.$$

Thus, the reconstruction error can be obtained

$$\left| \lim_{\delta \to 0} \frac{1}{\hat{g}(0)} \int_{|b+\varphi_k'(\omega)|<\tilde{\varepsilon}} S_{x,\tilde{\varepsilon}}^{\delta}(a,u) du - A_k(\omega) e^{i\varphi_k(\omega)} \right|$$

$$= \left| \left( \frac{1}{\hat{g}(0)} \int_{\{b:|b+\varphi_k'(\omega)|<\Delta\}} W_{e,x}^{\hat{g}}(a,b) db - \int_{\{b:|b+\varphi_k'(\omega)|<\Delta\}/A_{f,\tilde{\varepsilon}}(\omega)} W_{e,x}^{\hat{g}}(a,b) db \right) - A_k(\omega) e^{i\varphi_k(\omega)} \right|$$

$$\leq \left| \frac{1}{\hat{g}(0)} \int_{\{b:|b+\varphi_k'(\omega)|<\Delta\}} W_{e,x}^{\hat{g}}(a,b) db - A_k(\omega) e^{i\varphi_k(\omega)} \right| + \left| \frac{1}{\hat{g}(0)} \int_{\{b:|b+\varphi_k'(\omega)|<\Delta\}/A_{f,\tilde{\varepsilon}}(\omega)} W_{e,x}^{\hat{g}}(a,b) db \right|$$

$$\leq \left| \frac{1}{\hat{g}(0)} \int_{\{b:|b+\varphi_k'(\omega)|<\Delta\}} W_{e,x}^{\hat{g}}(a,b) - Q_{k,1}(a,b) db \right| + \left| \frac{1}{\hat{g}(0)} \int_{\{b:|b+\varphi_k'(\omega)|<\Delta\}} Q_{k,1}(a,b) db - A_k(\omega) e^{i\varphi_k(\omega)} \right|$$

$$+ \left| \frac{1}{\hat{g}(0)} \int_{\{b:|b+\varphi_k'(\omega)|<\Delta\}/A_{f,\tilde{\varepsilon}}(\omega)} W_{e,x}^{\hat{g}}(a,b) db \right|$$

$$\leq \frac{1}{\hat{g}(0)} \int_{\{b:|b+\varphi_k'(\omega)|<\Delta\}} \left| W_{e,x}^{\hat{g}}(a,b) - Q_{k,1}(a,b) \right| db + \frac{1}{\hat{g}(0)} \int_{\{b:|b+\varphi_k'(\omega)|<\Delta\}/A_{f,\tilde{\varepsilon}}(\omega)} \left| W_{e,x}^{\hat{g}}(a,b) \right| db$$

$$\leq \frac{1}{\hat{g}(0)} \int_{\{b:|b+\varphi_k'(\omega)|<\Delta\}} \varepsilon E_0(a) db + \frac{1}{\hat{g}(0)} \int_{\{b:|b+\varphi_k'(\omega)|<\Delta\}} \tilde{\varepsilon} db$$

$$\leq \frac{1}{\hat{g}(0)} \int_{\{b:|b+\varphi_k'(\omega)|<\Delta\}} 2\tilde{\varepsilon} db = \frac{4\tilde{\varepsilon}\Delta}{\hat{g}(0)}$$

Next, we will analyze the reconstruction of WTMSST. Suppose that all conditions in Definition 2 are satisfied. Select a function $\hat{h} \in C^{\infty}(\mathbb{R})$ with $\int \hat{h}(\omega) d\omega = 1$, and a window $\hat{g} \in S$ in Schwartz class such that its time-domain form $g$ is compactly supported in $[-\Delta, \Delta]$ where $\Delta = ad$. Moreover, $B_{x,\tilde{\varepsilon}}(\omega) := \{b \in \mathbb{R}; |W_{e,x}^{\hat{g}}(\omega_0/\omega,b)| > \tilde{\varepsilon}\}$ and the $\varepsilon$ is small enough. For all $\omega \in \mathbb{R}^+$, we assume that $\varepsilon \leq \Delta^3/8$. Then, for $\forall \omega \in \mathbb{R}^+$ and $\forall k \in \{1,\ldots,K\}$, we have

$$\left| \lim_{\delta \to 0} \frac{1}{\hat{g}(0)} \int_{|u+\varphi'(\omega)|<\tilde{\varepsilon}} S_{x,\tilde{\varepsilon}}^{[N],\delta}(a,u) du - A_k(\omega) e^{i\varphi_k(\omega)} \right| \leq C\tilde{\varepsilon}.$$

**Estimate 6.** For each component i.e., $k \in \{1,\ldots,K\}$, we can get

$$\left| A_k(\xi) - A_k(\omega) \right| \leq \varepsilon M'' |\xi - \omega|, \quad \left| \varphi_k''(\xi) - \varphi_k''(\omega) \right| \leq \varepsilon M'' |\xi - \omega| \text{ and } \left| \varphi_k'(\xi) - \varphi_k'(\omega) - \varphi_k''(\omega)(\xi - \omega) \right| \leq 0.5\varepsilon M'' |\xi - \omega|^2.$$

**Proof.** For any $\omega \geq 0$, we have

$$\left| A_k(\xi) - A_k(\omega) \right| = \left| \int_{\omega}^{\xi} A_k'(u) du \right| \leq \int_{\omega}^{\xi} \left| A_k'(u) \right| du \leq \varepsilon \int_{\omega}^{\xi} \left| \varphi_k''(u) \right| du = \varepsilon M'' |\xi - \omega|$$

$$\left| \varphi_k''(\xi) - \varphi_k''(\omega) \right| = \left| \int_{\omega}^{\xi} \varphi_k'''(u) du \right| \leq \varepsilon \int_{\omega}^{\xi} \left| \varphi_k''(u) \right| du = \varepsilon M'' |\xi - \omega|$$

$$\left|\varphi_k'(\xi)-\varphi_k'(\omega)-\varphi_k''(\omega)(\xi-\omega)\right|=\left|\int_\omega^\xi \varphi_k''(u)-\varphi_k''(\omega)du\right|\le 0.5\varepsilon M''|\xi-\omega|^2$$

**Estimate 7.** For $k\in\{1,\ldots,K\}$ and $(a,b)\in\mathbb{R}^+\times\mathbb{R}$, we have

$$\left|W_{e,x}^{\hat{g}}(a,b)-Q_{k,2}(a,b)\right|\le \varepsilon E_1(a)$$

where

$$E_1(a)=\frac{1}{2\pi}\left(\sum_{k=1}^K I_1\left|a^{-2}\right|M''+\sum_{k=1}^K \frac{1}{6}I_3\left|a^{-4}A_k(\omega)\right|M''\right)$$

with $I_n=\int |x|^n g(x)dx$ and $Q_{k,2}(a,b)$ represents the MWT of the signal with second-order GD, i.e.,

$$Q_{k,2}(a,b)=\sum_{k=1}^K \frac{1}{2\pi}\int_0^{+\infty} A_k(\omega)e^{i\left(\varphi_k(\omega)+\varphi_k'(\omega)(\xi-\omega)+0.5\varphi_k''(\omega)(\xi-\omega)^2\right)}\hat{g}(a(\xi-\omega))e^{i(\xi-\omega)b}d\xi. \tag{D2}$$

**Proof.** Combining Estimate 3 and D2 yields

$$\left|W_{e,x}^g(a,b)-Q_{k,2}(a,b)\right|$$

$$\le \frac{1}{2\pi}\sum_{k=1}^K \int_{\mathbb{R}^+}\left|A_k(\xi)-A_k(\omega)\right|\left|\hat{g}(a(\xi-\omega))\right|d\xi + \frac{1}{2\pi}\sum_{k=1}^K \int_{\mathbb{R}^+}\left|A_k(\omega)\right|\left|e^{i\left(\varphi_k(\xi)-\varphi_k(\omega)-\varphi_k'(\omega)(\xi-\omega)-0.5\varphi_k''(\omega)(\xi-\omega)^2\right)}-1\right|\left|\hat{g}(a(\xi-\omega))\right|d\xi$$

$$\le \frac{1}{2\pi}\sum_{k=1}^K \varepsilon\int_{\mathbb{R}^+} M''|\xi-\omega|\left|\hat{g}(a(\xi-\omega))\right|d\xi + \frac{1}{2\pi}\sum_{k=1}^K |A_k(\omega)|\int_{\mathbb{R}^+}\left|\varphi_k(\xi)-\varphi_k(\omega)-\varphi_k'(\omega)(\xi-\omega)-0.5\varphi_k''(\omega)(\xi-\omega)^2\right|\left|\hat{g}(a(\xi-\omega))\right|d\xi$$

$$\le \frac{1}{2\pi}\sum_{k=1}^K \varepsilon\int_{\mathbb{R}^+} M''|\xi-\omega|\left|\hat{g}(a(\xi-\omega))\right|d\xi + \frac{1}{2\pi}\sum_{k=1}^K |A_k(\omega)|\int_{\mathbb{R}^+}\left|\int_\omega^\xi\int_\omega^u \varphi_k''(x)-\varphi_k''(\omega)dxdu\right|\left|\hat{g}(a(\xi-\omega))\right|d\xi$$

$$\le \frac{1}{2\pi}\sum_{k=1}^K \varepsilon\int_{\mathbb{R}^+} M''|\xi-\omega|\left|\hat{g}(a(\xi-\omega))\right|d\xi + \frac{1}{2\pi}\sum_{k=1}^K |A_k(\omega)|\int_{\mathbb{R}^+}\left|\int_\omega^\xi\int_\omega^u \varphi_k''(x)-\varphi_k''(\omega)dxdu\right|\left|\hat{g}(a(\xi-\omega))\right|d\xi$$

$$\le \frac{1}{2\pi}\sum_{k=1}^K \varepsilon\int_{\mathbb{R}^+} M''|\xi-\omega|\left|\hat{g}(a(\xi-\omega))\right|d\xi + \frac{1}{2\pi}\sum_{k=1}^K \varepsilon|A_k(\omega)|\int_{\mathbb{R}^+}\frac{1}{6}M''|\xi-\omega|^3\left|\hat{g}(a(\xi-\omega))\right|d\xi$$

$$=\frac{\varepsilon}{2\pi}\left(\sum_{k=1}^K I_1\left|a^{-2}\right|M''+\sum_{k=1}^K \frac{1}{6}I_3\left|a^{-4}A_k(\omega)\right|M''\right)$$

If we introduce the same assumption as in Estimate 2, i.e., $\varepsilon\le E_1^{-2/3}(a)$, then we can obtain $\varepsilon E_0(a)\le \tilde{\varepsilon}$. Thus, for any $k\in\{1,\ldots,K\}$ and any $(a,b)\in Z_k$, we have $\left|W_{e,x}^{\hat{g}}(a,b)-Q_{k,2}(a,b)\right|\le \varepsilon E_1(a)\le \tilde{\varepsilon}$ and for any $(a,b)\notin Z_k$, we have $|W_{e,x}^g(a,b)|\le \varepsilon E_1(a)\le \tilde{\varepsilon}$

**Estimate 8.** Suppose that all the conditions in Estimate 5 hold, and for all $\omega\in\mathbb{R}^+$, we assume that $\varepsilon\le \Delta^3/8$. Further, for $\forall a\in\mathbb{R}^+$ and $\forall k\in\{1,\ldots,K\}$, we have

$$\left|\lim_{\delta\to 0}\frac{1}{\hat{g}(0)}\int_{|u+\varphi'(\omega)|<\tilde{\varepsilon}} S_{x,\tilde{\varepsilon}}^{[N],\delta}(a,u)du - A_k(\omega)e^{i\varphi_k(\omega)}\right|\le C\tilde{\varepsilon}.$$

**Proof.** According to Estimate 5, we can also obtain

$$\lim_{\delta\to 0}\frac{1}{\hat{g}(0)}\int_{|u+\varphi'(\omega)|<\tilde{\varepsilon}} S_{x,\tilde{\varepsilon}}^{[N],\delta}(a,u)du = \int_{A_{f,\tilde{\varepsilon}}(\omega)\cap\{b:|b+\varphi_k'(\omega)|<\Delta\}} W_{e,x}^{\hat{g}}(a,b)db$$

$$= \int_{\{b:|b+\varphi_k'(\omega)|<\Delta\}} W_{e,x}^{\hat{g}}(a,b)db - \int_{\{b:|b+\varphi_k'(\omega)|<\Delta\}/B_{x,\tilde{\varepsilon}}(\omega)} W_{e,x}^{\hat{g}}(a,b)db$$

According to the following relationship

$$\int_{\{b:|b+\varphi_k'(\omega)|<\Delta\}} Q_{k,1}(a,b)db = \sum_{k=0}^K \int_{\{b:|b+\varphi_k'(\omega)|<\Delta\}} \frac{1}{2\pi}\int_{-\infty}^{+\infty} A_k(\omega)e^{i\left(\varphi_k(\omega)+\varphi_k'(\omega)(\xi-\omega)+0.5\varphi_k''(\omega)(\xi-\omega)^2\right)}\hat{g}(a(\xi-\omega))e^{i(\xi-\omega)b}d\xi db$$

$$= \sum_{k=0}^K \frac{1}{2\pi}\int_{-\infty}^{+\infty} A_k(\omega)e^{i\left(\varphi_k(\omega)+\varphi_k'(\omega)(\xi-\omega)+0.5\varphi_k''(\omega)(\xi-\omega)^2\right)}\hat{g}(a(\xi-\omega))\int_{\{b:|b+\varphi_k'(\omega)|<\Delta\}} e^{i(\xi-\omega)b}dbd\xi$$

$$= \sum_{k=0}^K \hat{g}(0)A_k(\omega)e^{i\varphi_k(\omega)}.$$

we can obtain

$$\left|\lim_{\delta\to 0}\frac{1}{\hat{g}(0)}\int_{|b+\varphi'_k(\omega)|<\tilde{\varepsilon}}S_{x,\tilde{\varepsilon}}^{[N],\delta}(a,u)du-A_k(\omega)e^{i\varphi_k(\omega)}\right|$$

$$=\left|\left(\frac{1}{\hat{g}(0)}\int_{\{b:|b+\varphi'_k(\omega)|<\Delta\}}W_{e,x}^{\hat{g}}(a,b)db-\int_{\{b:|b+\varphi'_k(\omega)|<\Delta\}/B_{x,\tilde{\varepsilon}}(\omega)}W_{e,x}^{\hat{g}}(a,b)db\right)-A_k(\omega)e^{i\varphi_k(\omega)}\right|$$

$$\leq\left|\frac{1}{\hat{g}(0)}\int_{\{b:|b+\varphi'_k(\omega)|<\Delta\}}W_{e,x}^{\hat{g}}(a,b)-Q_{k,2}(a,b)db\right|+\left|\frac{1}{\hat{g}(0)}\int_{\{b:|b+\varphi'_k(\omega)|<\Delta\}}Q_{k,2}(a,b)db-A_k(\omega)e^{i\varphi_k(\omega)}\right|+\left|\frac{1}{\hat{g}(0)}\int_{\{b:|b+\varphi'_k(\omega)|<\Delta\}/B_{x,\tilde{\varepsilon}}(\omega)}W_{e,x}^{\hat{g}}(a,b)db\right|$$

$$\leq\frac{1}{\hat{g}(0)}\int_{\{b:|b+\varphi'_k(\omega)|<\Delta\}}\varepsilon E_1(a)db+\frac{1}{\hat{g}(0)}\int_{\{b:|b+\varphi'_k(\omega)|<\Delta\}}\tilde{\varepsilon}db$$

$$\leq\frac{1}{\hat{g}(0)}\int_{\{b:|b+\varphi'_k(\omega)|<\Delta\}}2\tilde{\varepsilon}db=\frac{4\tilde{\varepsilon}\Delta}{\hat{g}(0)}$$

It can be seen that WTMSST only improves the resolution of WTSST and does not destroy its reconstruction accuracy.

## APPENDIX E

### DERIVATION OF ITERATION SPEED

Linear iterative implementation of discrete algorithm

$$\hat{t}^{[1]}(a,b)=-\varphi'(\omega)+\frac{\varphi''(\omega)^2}{\varphi''(\omega)^2+(a^2\sigma)^2}(\varphi'(\omega)+b)$$

$$\hat{t}^{[2]}(a,b)=\hat{t}^{[1]}(a,\hat{t}^{[1]}(a,b))=-\varphi'(\omega)+\left[\frac{\varphi''(\omega)^2}{\varphi''(\omega)^2+(a^2\sigma)^2}\right]^2(\varphi'(\omega)+b)$$

$$\hat{t}^{[3]}(a,b)=\hat{t}^{[1]}(a,\hat{t}^{[1]}(a,\hat{t}^{[1]}(a,b)))=-\varphi'(\omega)+\left[\frac{\varphi''(\omega)^2}{\varphi''(\omega)^2+(a^2\sigma)^2}\right]^3(\varphi'(\omega)+b) \quad (E1)$$

$$\vdots$$

$$\hat{t}^{[N]}(a,b)=\hat{t}^{[1]}(a,\hat{t}^{[N-1]}(a,b))=-\varphi'(\omega)+\left[\frac{\varphi''(\omega)^2}{\varphi''(\omega)^2+(a^2\sigma)^2}\right]^N(\varphi'(\omega)+b)$$

Equation E1 shows that it needs $N$ iterations to obtain $\hat{t}^{[N]}(a,b)$.

Nonlinear iterative implementation of discrete algorithm

$$\hat{t}^{[1]}(a,b)=-\varphi'(\omega)+\frac{\varphi''(\omega)^2}{\varphi''(\omega)^2+(a^2\sigma)^2}(\varphi'(\omega)+b)$$

$$\hat{t}^{[2]}(a,b)=\hat{t}^{[1]}(a,\hat{t}^{[1]}(a,b))=-\varphi'(\omega)+\left[\frac{\varphi''(\omega)^2}{\varphi''(\omega)^2+(a^2\sigma)^2}\right]^2(\varphi'(\omega)+b)$$

$$\hat{t}^{[4]}(a,b)=\hat{t}^{[2]}(a,\hat{t}^{[2]}(a,b))=-\varphi'(\omega)+\left[\frac{\varphi''(\omega)^2}{\varphi''(\omega)^2+(a^2\sigma)^2}\right]^4(\varphi'(\omega)+b)$$

$$\hat{t}^{[8]}(a,b)=\hat{t}^{[4]}(a,\hat{t}^{[4]}(a,b))=-\varphi'(\omega)+\left[\frac{\varphi''(\omega)^2}{\varphi''(\omega)^2+(a^2\sigma)^2}\right]^8(\varphi'(\omega)+b) \quad (E2)$$

$$\hat{t}^{[16]}(a,b)=\hat{t}^{[8]}(a,\hat{t}^{[8]}(a,b))=-\varphi'(\omega)+\left[\frac{\varphi''(\omega)^2}{\varphi''(\omega)^2+(a^2\sigma)^2}\right]^{16}(\varphi'(\omega)+b)$$

$$\vdots$$

$$\hat{t}^{[N]}(a,b)=\hat{t}^{[N/2]}(a,\hat{t}^{[N/2]}(a,b))=-\varphi'(\omega)+\left[\frac{\varphi''(\omega)^2}{\varphi''(\omega)^2+(a^2\sigma)^2}\right]^N(\varphi'(\omega)+b)$$

It can be seen that E2 only needs $\log_2 N$ iterations to obtain the same result as the N iterations in E1.

## ACKNOWLEDGMENT

The authors are grateful for the valuable comments from the associate editor and reviewers to help greatly refine the quality of the paper.